\documentclass[aps,prx,preprint,onecolumn,citeautoscript,footinbib,
eqsecnum]{revtex4-1}  
\usepackage{amsfonts,amsthm,amsmath,amssymb,upgreek}
\usepackage{amsmath,amssymb,bm} 
\usepackage{graphicx}  
\usepackage[export]{adjustbox}
\usepackage{color} 
\usepackage[dvipsnames]{xcolor}
\usepackage[papersize={8.5in,11in}]{geometry}
\usepackage[colorlinks=true]{hyperref}  
\usepackage{ulem}
\usepackage[mathscr]{euscript}
\usepackage[section]{placeins}
\hypersetup{
    bookmarks=true,         
    unicode=false,          
    pdftoolbar=true,        
    pdfmenubar=true,        
    pdffitwindow=false,     
    pdfstartview={FitH},    
    pdfsubject={},   
    pdfcreator={},   
    pdfproducer={}, 
    pdfkeywords={} {} {}, 
    pdfnewwindow=true,      
    colorlinks=true,       
    linkcolor=magenta, 
    citecolor=blue,        
    filecolor=magenta,      
    urlcolor=blue           
} 

\usepackage{subfig}

\usepackage{amsfonts}
\usepackage{upgreek}
\usepackage{slashed}
\usepackage{latexsym}

\newcommand{\change}{\color{black}}

\newcommand{\beq}{\begin{equation}}
\newcommand{\eeq}{\end{equation}}
\def\bea{\begin{eqnarray}}
\def\eea{\end{eqnarray}}

\renewcommand{\vec}[1]{\boldsymbol{#1}}

\newcommand{\lam}{\lambda} 
\newcommand{\kin}{s_0} 
\newcommand{\gc}{g_0} 
\newcommand{\gcr}{g} 
\newcommand{\gam}{\gamma_0} 
\newcommand{\gamr}{\gamma} 
\newcommand{\gv}{v_{0}} 
\newcommand{\gvr}{v} 

\newcommand{\fd}[1]{f^{\dagger}_{#1}} 
\newcommand{\fa}[1]{f_{#1}} 
\newcommand{\bd}{b^{\dagger}} 
\newcommand{\ba}{b} 
\newcommand{\cd}[1]{\psi^{\dagger}_{#1}} 
\newcommand{\ca}[1]{\psi_{#1}} 
\newcommand{\pa}{\phi_{a}} 
\newcommand{\zt}{\zeta} 
\newcommand{\No}{N_{0}} 
\newcommand{\LS}{\Lambda_{S}}  
\newcommand{\LP}{\Lambda_{c}}
\newcommand{\LN}{\Lambda_{n}}

\newcommand{\al}{\alpha}
\newcommand{\be}{\beta}

\newcommand{\up}{^}

\newcommand{\betg}{\beta (g)} 
\newcommand{\betgam}{\beta (\gamma)} 
\newcommand{\bets}{\beta (s)} 
\newcommand{\betv}{\beta (v)} 

\newcommand{\rb}{\bar{r}}
\newcommand{\ep}{\epsilon}
\newcommand{\epp}{\epsilon'}

\newcommand{\zf}{Z_{f}}
\newcommand{\zb}{Z_{b}}
\newcommand{\zg}{Z_{g}}

\newcommand{\zgam}{Z_{\gamma}}
\newcommand{\zv}{Z_{v}}

\newcommand{\rgs}{\mu}

\newcommand{\iw}{i\omega}
\newcommand{\inu}{i\nu}

\newcommand{\lo}{L_{0}}
\newcommand{\lop}{L'_{0}}
\newcommand{\lopp}{L''_{0}}
\newcommand{\loppp}{L'''_{0}}

\newcommand{\Lgam}{L_{\gamma}}
\newcommand{\Lg}{L_{g}}
\newcommand{\Lv}{L_{v}}

\newcommand{\la}{L_{1}}
\newcommand{\lap}{L'_{1}}
\newcommand{\lapp}{L''_{1}}
\newcommand{\lappp}{L'''_{1}}

\newcommand{\lb}{L_{2}}
\newcommand{\lbp}{L'_{2}}
\newcommand{\lbpp}{L''_{2}}
\newcommand{\lbppp}{L'''_{2}}

\newcommand{\lc}{L_{3}}
\newcommand{\lcp}{L'_{3}}
\newcommand{\lcpp}{L''_{3}}
\newcommand{\lcppp}{L'''_{3}}

\newcommand{\Da}{D_{1\phi}}
\newcommand{\Db}{D_{2\phi}}
\newcommand{\Dc}{D_{3\phi}}

\newcommand{\Dap}{D'_{1\psi}}
\newcommand{\Dbp}{D'_{2\psi}}
\newcommand{\Dcp}{D'_{3\psi}}

\newcommand{\Dapp}{D''_{1\psi}}
\newcommand{\Dbpp}{D''_{2\psi}}
\newcommand{\Dcpp}{D''_{3\psi}}

\newcommand{\Daz}{D_{1\zeta}}
\newcommand{\Dbz}{D_{2\zeta}}
\newcommand{\Dcz}{D_{3\zeta}}

\newcommand{\po}{P_{0}}
\newcommand{\Pgam}{P_{\gamma}}
\newcommand{\Pg}{P_{g}}
\newcommand{\Pv}{P_{v}}

\newcommand{\Pa}{P_{1}}
\newcommand{\pap}{P'_{1}}
\newcommand{\papp}{P''_{1}}
\newcommand{\pappp}{P'''_{1}}

\newcommand{\pb}{P_{2}}
\newcommand{\pbp}{P'_{2}}
\newcommand{\pbpp}{P''_{2}}
\newcommand{\pbppp}{P'''_{2}}

\newcommand{\pc}{P_{3}}
\newcommand{\pcp}{P'_{3}}
\newcommand{\pcpp}{P''_{3}}
\newcommand{\pcppp}{P'''_{3}}

\newcommand{\To}{T_{0}}
\newcommand{\Tgam}{T_{\gamma}}
\newcommand{\Tg}{T_{g}}
\newcommand{\Tv}{T_{v}}

\newcommand{\Ta}{T_{1}}
\newcommand{\Tap}{T'_{1}}
\newcommand{\Tapp}{T''_{1}}
\newcommand{\Tappp}{T'''_{1}}

\newcommand{\Tb}{T_{2}}
\newcommand{\Tbp}{T'_{2}}
\newcommand{\Tbpp}{T''_{2}}
\newcommand{\Tbppp}{T'''_{2}}

\newcommand{\Tc}{T_{3}}
\newcommand{\Tcp}{T'_{3}}
\newcommand{\Tcpp}{T''_{3}}
\newcommand{\Tcppp}{T'''_{3}}

\newcommand{\nt}{\tilde{n}}
\newcommand{\nf}{n_{f}}

\newcommand{\lbr}{\langle}
\newcommand{\rbr}{\rangle}

\begin{document}


\title{Anomalous density fluctuations in a random $t$-$J$ model}

\author{Darshan G. Joshi}
\affiliation{Department of Physics, Harvard University, Cambridge MA 02138, USA}

\author{Subir Sachdev}
\affiliation{Department of Physics, Harvard University, Cambridge MA 02138, USA}

\date{\today
\\
\vspace{0.4in}}

\begin{abstract}
A previous work (Joshi {\it et al.\/}, Phys. Rev. X {\bf 10}, 021033 (2020)) found a deconfined critical point at non-zero doping in a $t$-$J$ model with all-to-all and random hopping and spin exchange, and argued for its relevance to the phenomenology of the cuprates. We extend this model to include all-to-all and random density-density interactions of mean-square strength $K$. 
In a fixed realization of the disorder, and for specific values of the hopping, exchange, and density interactions, the model is supersymmetric; but, we find no supersymmetry after independent averages over the interactions. 
Using the previously developed renormalization group analysis, we find a new fixed point at non-zero $K$. However, this fixed point is unstable towards the previously found fixed point at $K=0$ in our perturbative analysis. We compute the exponent characterizing density fluctuations at both fixed points: this exponent determines the spectrum of electron energy-loss spectroscopy.
\end{abstract}

\maketitle


\section{Introduction}
\label{sec:intro}

The possibility of a quantum critical point underneath the superconducting dome of high-temperature cuprate materials has been a subject of intense study. Photoemission experiments \cite{Shen18, Shen19} and thermal Hall measurements \cite{Michon18} have given strong evidence for a transformation in the Fermi surface across a critical value of doping. Such a critical point, and the corresponding critical theory, possibly holds the key to understanding the enigmatic strange-metal phase at high temperatures.  The strange-metal phase is also characterized by an absence of quasiparticles and thus one expects a continuum response to many probes. {\change It is challenging to investigate the strange metal region with high resolution measurements, but remarkable progress has been made in this direction in the last few years.} Recently, an anomalous continuum was observed in dynamic charge response measurements \cite{Mitrano18, Husain19} on optimally doped Bi$_{2.1}$Sr$_{1.9}$Ca$_{1.0}$Cu$_{2.0}$O$_{8+x}$ (Bi-2212) using momentum-resolved electron energy-loss spectroscopy (M-EELS). {\change The dynamic charge response is directly related to the imaginary part of density-density correlation. Similar measurements have also revealed surprising results in the case of Sr$_{2}$RuO$_{4}$ \cite{Husain20}. These interesting set of experiments call for a quantitative theoretical investigation of the density-density correlation. }

{\change Along with collaborators}, we have recently proposed a microscopic model which hosts {\change a finite doping } quantum critical point \cite{dqcp_tj}. It was shown to be a deconfined critical point with a SYK-like \cite{SY92, kitaev2015talk} local spin correlations, {\change {\it i.e.\/},  $\langle \vec{S}(\tau) \cdot \vec{S}(0)\rangle \sim 1/|\tau|$, where $\tau$ is imaginary time}. {\change The model considered in Ref. \cite{dqcp_tj} has random and all-to-all hopping and exchange interactions, and was solved using a perturbative RG which yielded some exponents to all orders. In this work, we extend the model in Ref.~\cite{dqcp_tj} to include random and all-to-all density-density interactions. 
Motivated by the above mentioned M-EELS measurements, we will also compute the density-density correlation function in the model of Ref.~\cite{dqcp_tj}, and in the extended model. We find critical density-density correlations characterized by an exponent $\eta_n$, as specified by Eqs.~(\ref{chin}-\ref{chin3}) in the concluding Section~\ref{sec:conc}. A disordered Fermi liquid has $\eta_n = 2$, while the `marginal' value $\eta_n =1$ is observed in the M-EELS experiments, showing a striking non-Fermi liquid behavior with an anomalous enhancement of local density flucutations. 
We will find a new fixed point in the extended model where we establish that $\eta_n = 1$ to all orders in the perturbative RG.  To our knowledge, such a density correlation has not been quantitatively calculated in a microscopic model before, especially at a finite doping quantum critical point.

As we will discuss in detail below, our perturbative RG finds that the new fixed point is multi-critical, and unstable towards the fixed point found earlier in Ref.~\cite{dqcp_tj}. However, it could well be that this is a feature of the one-loop RG, and that, at higher orders, the new fixed point is a conventional critical point requiring only one tuning parameter. 
We will also compute the value $\eta_n$ at the fixed point of Ref.~\cite{dqcp_tj}, although we are only able to do this at the one loop level.}

The paper is organized as follows. In Sec. \ref{sec:model} we describe our model and related algebra of the operators. In Sec. \ref{sec:imp_H} we discuss the mapping of our model to an impurity model, which can be then studied using renormalization group as shown in Sec. \ref{sec:rg_fb}. 
{\change In this section we also present the main result of our work, {\it i.e.\/}, the exponent $\eta_n$ corresponding to the density correlator, which characterizes the anomalous density fluctuation. The RG analysis is performed at one-loop order.} We conclude in Sec. \ref{sec:conc} and present an alternative RG calculation in Appendix \ref{sec:rg_scn}. A discussion on possibility of supersymmetry can be found in Appendix \ref{app:bath}.

\section{Model}
\label{sec:model}

We consider the following Hamiltonian,
\begin{equation}
H_{tJK} = \frac{1}{\sqrt{N}} \sum_{ij} t_{ij} c\up{\dagger}_{i \al} c_{j \al} 
+ \frac{1}{\sqrt{N}} \sum_{i<j} J_{ij} \vec{S}_{i}\cdot\vec{S}_{j} + \frac{1}{\sqrt{N}} \sum_{i<j} K_{ij} \frac{n_{i} n_{j}}{4}  - \mu \sum_i c\up{\dagger}_{i \al} c_{i \al}  \,,
\label{eq:ham_tJK}
\end{equation}
where $N$ is the number of sites, $\mu$ is the chemical potential, $\alpha$ is the spin index ($\uparrow$ or $\downarrow$), $n_{i}=c\up{\dagger}_{i \al} c_{i \al}$ and double occupancy on each site is excluded, {\it i.e.\/}, $n_{i}\leq 1$. The {\change complex} hoppings $t_{ij}$, {\change real} exchange interactions $J_{ij}$, and {\change real} density-density interactions $K_{ij}$ are random numbers drawn from a Gaussian probability distribution with zero mean value such that $\overline{|t_{ij}|\up{2}}=t\up{2}$, $\overline{|J_{ij}|\up{2}}=J\up{2}$ and $\overline{|K_{ij}|\up{2}}=K\up{2}$. Note that the density-density interactions are present in the familiar derivation of the $t$-$J$ model from the Hubbard model, and are usually ignored. We include them here as independent random couplings, because we are interested in their possible influence on the spectrum of density fluctuations. 

To account for the double occupancy constraint, we  fractionalize the electron on each site into a bosonic holon ($b$) and fermionic spinon ($\fa{\al}$) degrees of freedom such that, 
\begin{equation}
c_{\al} = \fa{\al} \bd \,, ~~~ S\up{a} = \fd{\al}\frac{\sigma\up{a}_{\al\be}}{2} \fa{\be} \,, ~~~
V = \frac{1}{2}\fd{\al}\fa{\al} + \bd \ba \,, ~~~
n = \fd{\al}\fa{\al} \,. \label{fractionalize}
\end{equation}
The Hilbert-space constraint of no double occupancy now takes the form: $\fd{\al}\fa{\al} + \bd\ba =1$. Note that $V_{i} = 1- n_{i}/2$. 

On each site $i$, the operators $c$, $S$ and $V$ (dropping site indices) define a superalgebra $SU(1|2)$ as follows:
\begin{align}
\label{super}
\{c_\alpha , c_\beta \} &= 0 \,, ~~~ 
\{c_\alpha , c_\beta^\dagger\} = \delta_{\alpha\beta} V +  \sigma^a_{\alpha \beta} S^a \,, ~~~
[ S^a , c_\alpha ] = - \frac{1}{2} \sigma^a_{\alpha\beta} c_\beta \,, ~~~
[ S^a , c_\alpha^\dagger ] = \frac{1}{2} \sigma^a_{\beta\alpha} c_\beta^\dagger \,, \nonumber \\
[S^a, S^b] &= i \epsilon_{abc} S^c \,, ~~~
[S^a , V] = 0 \,, ~~~
[V, c_\alpha ] = \frac{1}{2} c_\alpha \,, ~~~ 
[V, c_\alpha^\dagger] = -\frac{1}{2} c_\alpha^\dagger \,. \end{align}
As an aside, note that one can also work with an alternative equivalent representation with a bosonic spinon and fermionic holon, which form a $SU(2|1)$ superalgebra \cite{dqcp_tj}.

The Hamiltonian $H_{tJK}$ clearly commutes with total spin, $\sum_i S^a_i$, and total density $\sum_i V_i$.
For the remaining generator, {\change $\sum_i c_{i \alpha}$}, of the $SU(1|2)$ superalgebra, the commutator is simple for 
for $t_{ij}=K_{ij}/2=-J_{ij}/2$, when we find
\beq
\left[ \sum_i c_{i \alpha}, H_{tJK} \right] = - \mu \, \sum_i c_{i \alpha}\,.
\label{cHtJK}
\eeq
{\change which connects the energy eigenvalues at different particle number}.
The non-random supersymmetric $t-J$ model has been studied in the past in one dimension, for instance see Refs. \cite{Wiegmann88, Forster89, Blatter90, Czech03, Sarkar91, Essler92}.  


\section{Large-$N$ limit and Impurity Hamiltonian}
\label{sec:imp_H}

{\change We can now make progress by resorting to the replica trick and taking the large volume limit, $N \rightarrow \infty$. Within this approach one first introduces field replicas, and the random coupling constants (here $t_{ij}$, $J_{ij}$ and $K_{ij}$) are averaged over. In many situations, such as in the spin-glass phase, the replica structure plays an important role. However, in our case we will be working at criticality, and we do not expect the replica structure to play a significant role. Therefore we do not write the replica indices in the subsequent discussion. Now taking the large volume limit we obtain the following single-site action:
}
\begin{eqnarray}
\mathcal{Z} &=& \int \mathcal{D} c_\alpha (\tau) e^{-\mathcal{S}-\mathcal{S_\infty}} \nonumber \\
\mathcal{S} &=& \int d \tau \left[c_\alpha^\dagger (\tau)  \left(\frac{\partial}{\partial \tau}-\mu \right)  c_\alpha (\tau) \right] 
 + t^2 \int d\tau d \tau' R (\tau - \tau') c_\alpha^\dagger (\tau) c_\alpha (\tau')  \nonumber \\
&~&~~ - \frac{J^2}{2} \int d\tau d \tau' Q (\tau - \tau') \vec{S} (\tau) \cdot \vec{S} (\tau') 
- \frac{K^2}{2} \int d\tau d \tau' P (\tau - \tau') n (\tau) n (\tau')
\,, \label{Zc}
\end{eqnarray}
where the fields $R$, $Q$, and $P$ have to be determined self-consistently via,
\begin{equation}
\label{eq:self_cons}
R(\tau - \tau') = -  \left\langle c^{}_{\alpha} (\tau) c^{\dagger}_\alpha (\tau') \right\rangle_\mathcal{Z} \,, ~~
Q (\tau - \tau') =  \frac{1}{3} \left\langle \vec{S} (\tau) \cdot \vec{S} (\tau') \right\rangle_\mathcal{Z} \,, ~~
P (\tau - \tau') =  \left\langle n (\tau) n (\tau') \right\rangle_\mathcal{Z} \,.
\end{equation}
{\change Here $\langle \dots \rangle_{\mathcal{Z}}$ means expectation value with respect to the partition function defined in Eq. (\ref{Zc}). }

To set-up our RG, let us ignore the self-consistency for now. We shall come back to it later. Let us assume that at the criticality the fields have the following power-law decay in imaginary time:
\begin{equation}
P(\tau) \sim \frac{1}{|\tau|^{d'-1}} \quad, \quad
Q(\tau) \sim \frac{1}{|\tau|^{d-1}} \quad, \quad 
R(\tau) \sim \frac{\mbox{sgn}(\tau)}{|\tau|^{r+1}}\,. 
\label{QRpower}
\end{equation}
Now we introduce fermionic and bosonic fields in the same spirit as in Ref. \cite{dqcp_tj} in order to obtain an impurity Hamiltonian. Such an impurity {\change action} has been studied in different limits in Refs. \cite{SBV1999,VBS2000,SS2001,VojtaFritz04,FritzVojta04,FritzThesis,Si1993a,Si1993b}. In our case we can map the above Hamiltonian to the following impurity and bath Hamiltonians:
\begin{align}
\label{eq:H_imp}
H_{\rm imp} &= (\kin + \lam)\fd{\al}\fa{\al} + \lam\bd\ba + \gc \left( \fd{\al}\ba\ca{\al}(0) + H.c.\right) 
+ \gam \fd{\al} \frac{\sigma\up{a}_{\al \be}}{2} \fa{\be} \pa(0) + \gv (\fd{\al}\fa{\al} - n_{f} )\zt(0) \nonumber \\
H_{\rm bath} &= \int |k|\up{r} dk \, k \, \cd{k\al}\ca{k\al} + \frac{1}{2} \int d\up{d}x \left( \pi_{a}\up{2} + (\partial_{x}\pa)\up{2} \right) 
+ \frac{1}{2} \int d\up{d'}x \left( \tilde{\pi}\up{2} + (\partial_{x}\zt)\up{2} \right) \,,
\end{align}
where $\lam \rightarrow \infty$ is introduced to handle the constraint $\fd{\al} \fa{\al} + \bd \ba = 1$, and $n_f = \left\langle f_\alpha^\dagger f_\alpha \right\rangle$. We have introduced fermionic bath $\ca{k\al}$, as well as bosonic baths $\pa$ and $\zt$, which upon integrating out gives us the original Hamiltonian. Also, $\pa (0) \equiv \pa (x=0)$, $\zt (0) \equiv \zt (x=0)$ and $\ca{\al} (0) \equiv \int dk |k|\up{r} \ca{k\al}$.

The Hamiltonian $H_{\rm imp}+H_{\rm bath}$ is our representation of the effective theory after averaging the disorder. We explore the possibility that this Hamiltonian could be supersymmetric in Appendix~\ref{app:bath}, and find no supersymmetry. So supersymmetry is specific to particular realizations of disorder, and does not re-emerge after independent averages over $t_{ij}$, $J_{ij}$, and $K_{ij}$.
{\change Perhaps if we begin strictly with the condition of supersymmetry for each disorder realization (i.e. $t_{ij}=K_{ij}/2=-J_{ij}/2$) then the disorder average might be supersymmetric. However, this means that there is only one independent random variable. This brings along difficultly when doing disorder average since it will result in several cross-terms like $S(\tau) n(\tau')$ etc. We have avoided this complication here. Another route may be to choose the distribution of random variables such that their means have the ratios required by supersymmetry. However, this goes beyond the scope of present work and we have not explored this possibility. 
}

\section{Renormalization group analysis}
\label{sec:rg_fb}

In this section we present the details of RG analysis of the impurity Hamiltonian introduced in Eq. \ref{eq:H_imp}. At the tree-level the scaling dimensions are found as follows:
\begin{align}
&\text{dim} [\fa{}] = \text{dim} [\ba] = 0 \,, ~~~ \text{dim} [\ca{k\al}] = -\frac{1+r}{2} = -\text{dim} [\ca{\al} (0)]   \,, ~~~ \text{dim} [\pa] = \frac{d-1}{2} \,, ~~~ 
\text{dim} [\zt] = \frac{d'-1}{2} \,
\nonumber \\
&\text{dim} [\gc] = \frac{1-r}{2} \equiv \rb \,, ~~~ \text{dim} [\gam] = \frac{3-d}{2} \equiv \frac{\ep}{2} \,, 
~~~ \text{dim} [\gv] = \frac{3-d'}{2} \equiv \frac{\epp}{2} \,.
\label{eq:sca_dimf}
\end{align}
This establishes $r=1$, $d=3$, and $d'=3$ as upper critical dimensions.
Next, the renormalized fields and couplings are defined as follows:
\begin{equation}
\fa{\al} = \sqrt{\zf} \fa{R \al} \,, ~~ \ba = \sqrt{\zb} \ba_{R} \,, ~~ 
\gc = \frac{\rgs\up{\rb} \zg}{\sqrt{\zf \zb}} \gcr \,,~~ 
\gam = \frac{\rgs\up{\ep/2} \zgam}{\zf \sqrt{\tilde{S}_{d+1}}} \gamr \,, ~~ 
\gv = \frac{\rgs\up{\epp/2} \zv}{\zf \sqrt{\tilde{S}_{d'+1}}} \gvr \,,
\label{eq:renorm_factf}
\end{equation}
where $\tilde{S}_d = \Gamma (d/2-1)/(4 \pi\up{d/2})$. The bulk-bath fields $\ca{}$, $\pa$, and $\zt$ do not get renormalized because of the absence of the respective interaction terms.
These renormalization factors, {\change $Z's$,} will be determined in the following sections from the self-energy and vertex corrections. We shall work at zero temperature and tune the system to criticality, i.e., we set $\kin=0$ and subsequently derive the flow away from it.


\subsection{Self energy}
\label{sec:se_f}

\begin{figure}[t]
\centering
\includegraphics[width=0.8\textwidth]{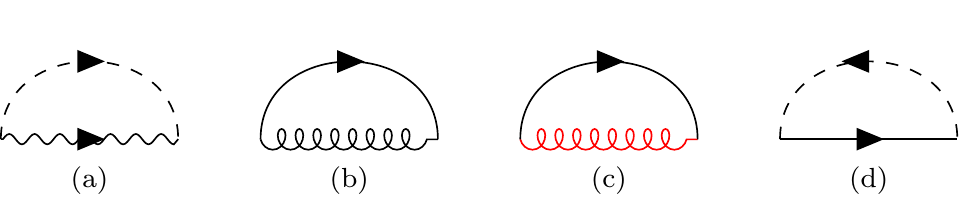}
\caption{One-loop fermion and boson self-energy diagrams. Fermion self-energy diagrams are shown in (a), (b), and (c), while boson self energy is shown in (d). We use a convention where a solid line denotes $f$ propagator, a dashed line denotes $\psi$ propagator, wavy denotes $b$ propagator, spiral denotes $\phi$ propagator, and red spiral denotes $\zeta$ propagator.}
\label{fig:dia_sef}
\end{figure}

We begin with the calculation of the fermionic self energy at one-loop level. Note that at this level there are no diagrams involving both the bosonic and the fermionic bath couplings. Here we have three relevant diagrams, shown in Fig. \ref{fig:dia_sef} (a), (b) and (c). The diagrams in Fig. \ref{fig:dia_sef} (a) and (b) have been evaluated already, and their corresponding expressions can be found in Eqs. (3.3) and (3.4) in  Ref. \cite{dqcp_tj}, respectively.  Below we quote the fermion self-energy corresponding to the diagram in Fig. \ref{fig:dia_sef} (c), 
\begin{align}
\label{eq:se_f3f}
\Sigma\up{f}_{\ref{fig:dia_sef}(c)} &= 
\gv\up{2} \frac{1}{\beta} \sum_{i\omega_{n}} \int \frac{d\up{d'}k}{(2\pi)\up{d'}} \frac{1}{\omega_{n}\up{2} +k\up{2}} 
\frac{1}{\inu + \iw - \lam} 
= \gv\up{2} \frac{S_{d'}}{2} \int_{0}\up{\infty} dk \frac{k\up{d'-2} }{\inu -\lam -k} \nonumber \\
&= \gv\up{2} \frac{S_{d'}}{2} \pi  \csc (\pi (d'-2)) (\lam-i \nu )^{-2+d'}
\nonumber \\
&= C_{\mu} \gvr\up{2} (\inu-\lam) 
\left[-\frac{1}{\epp} + \frac{1}{2} (\No + 2 i \pi ) \right]  ~~~
(\text{with $C_{\mu} = \mu\up{\epp} (\inu-\lam)\up{-\epp} \frac{\zv\up{2}}{\zb\up{2}}$}) \,.
\end{align}
Here, $\No=\gamma_{E} -2 \log (2)-\psi ^{(0)}\left(\frac{3}{2}\right)$ with $\gamma_{E}$ being the Euler's constant and $\psi ^{(0)}$ is the polygamma function.



There is only one diagram contributing to the bosonic self-energy at one-loop level, shown in Fig. \ref{fig:dia_sef} (d). 
It has been evaluated previously and its expression can be found in Eq. (3.8) in Ref. \cite{dqcp_tj}.

\begin{figure}[t]
\centering
\includegraphics[width=0.9\textwidth]{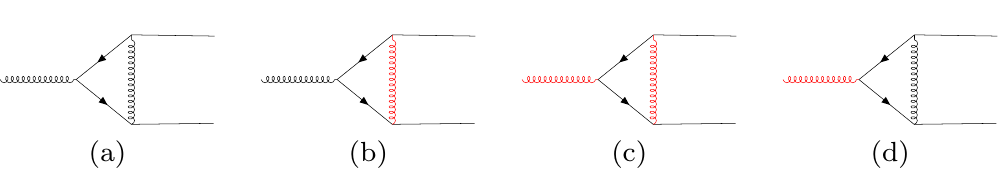} 
\caption{One-loop diagrams for vertex corrections. Vertex corrections to $\gam$ are shown in (a) and (b), while that for $\gv$ are shown in (c) and (d).
The convention for different lines is same as introduced in Fig. \ref{fig:dia_sef}.}
\label{fig:dia_vf}
\end{figure}

\subsection{Vertex correction}
\label{sec:ver}

Firstly, note that there is no one-loop correction to the vertex $\gc$ corresponding to the fermionic bath coupling. So we proceed with calculating the vertex corrections to the bosonic bath couplings $\gam$ and $\gv$. The diagrams corresponding to the vertex correction to $\gam$ are shown in Fig. \ref{fig:dia_vf} (a) and (b), while those corresponding to $\gv$ are shown in Fig. \ref{fig:dia_vf} (c) and (d). Note that the diagram in Fig. \ref{fig:dia_vf} (a) has been evaluated before and its expression can be found in Eq. (3.9) in Ref. \cite{dqcp_tj}. The expressions for the rest of the diagrams in Fig. \ref{fig:dia_vf} are as follows:
\begingroup
\allowdisplaybreaks
\begin{align}
\label{eq:vertex_gam2}
\Gamma\up{\gamma}_{\ref{fig:dia_vf} (b)} &= \gam \gv\up{2} \frac{1}{\beta} \sum_{i\omega_{1n}} \int d\up{d'}k_{1} 
\frac{1}{\omega_{1n}\up{2} + k_{1}\up{2}} \frac{1}{i\Omega_{1n}+i\omega_{1n}-\lam} \frac{1}{i\Omega_{2n}+i\omega_{1n}-\lam} 
\nonumber \\
&= \gam \gv\up{2} \int \frac{d\up{d'}k_{1} }{2k_{1}} 
\frac{1}{i\Omega_{1n}-k_{1}-\lam} \frac{1}{i\Omega_{2n}-k_{1}-\lam} 
= \gam C_{\mu} \gvr\up{2} 
\left[ \frac{1}{\epp} -1 + \frac{1}{2} \left(-\No -2 i \pi \right) \right] \,, \\
\label{eq:vertex_gv1}
\Gamma\up{\gvr}_{\ref{fig:dia_vf} (c)} &= \gv\up{3} \frac{1}{\beta} \sum_{i\omega_{1n}} \int d\up{d'}k_{1} 
\frac{1}{\omega_{1n}\up{2} + k_{1}\up{2}} \frac{1}{i\Omega_{1n}+i\omega_{1n}-\lam} \frac{1}{i\Omega_{2n}+i\omega_{1n}-\lam} 
\nonumber \\
&= \gam\up{3} \int \frac{d\up{d'}k_{1} }{2k_{1}} 
\frac{1}{i\Omega_{1n}-k_{1}-\lam} \frac{1}{i\Omega_{2n}-k_{1}-\lam} 
=\gv C_{\mu} \gvr\up{2} 
\left[ \frac{1}{\epp} -1 + \frac{1}{2} \left(-\No -2 i \pi \right) \right] \,, \\
\label{eq:vertex_gv2}
\Gamma\up{\gvr}_{\ref{fig:dia_vf} (d)} &= \frac{3}{4} \gv \gam\up{2} \frac{1}{\beta} \sum_{i\omega_{1n}} \int d\up{d}k_{1} 
\frac{1}{\omega_{1n}\up{2} + k_{1}\up{2}} \frac{1}{i\Omega_{1n}+i\omega_{1n}-\lam} \frac{1}{i\Omega_{2n}+i\omega_{1n}-\lam} 
\nonumber \\
&= \frac{3}{4} \gv \gam\up{2} \int \frac{d\up{d}k_{1} }{2k_{1}} 
\frac{1}{i\Omega_{1n}-k_{1}-\lam} \frac{1}{i\Omega_{2n}-k_{1}-\lam} 
=\frac{3}{4} \gv B_{\mu} \gamr\up{2} 
\left[ \frac{1}{\ep} -1 + \frac{1}{2} \left(-\No -2 i \pi \right) \right] \,.
\end{align}
\endgroup


\subsection{Beta functions}
\label{sec:beta}

In the expressions for the renormalized vertices and the $f/b$ Green's functions, we look at the cancellation of poles at the external frequency $\inu -\lam=\rgs$. We thus obtain the  following expressions of the renormalizing factors,
\begin{align}
\label{eq:zff}
\zf &= 1 - \frac{\gcr\up{2}}{2\rb} - \frac{3\gamr\up{2}}{4\ep} - \frac{\gvr\up{2}}{\epp} \,, \\
\label{eq:zbf}
\zb &= 1 - \frac{\gcr\up{2}}{\rb}  \,, \\
\label{eq:zgamf}
\zgam &= 1 + \frac{\gamr\up{2}}{4\ep} - \frac{\gvr\up{2}}{\epp} \,, \\
\label{eq:zvf}
\zv &= 1 - \frac{\gvr\up{2}}{\epp} - \frac{3\gamr\up{2}}{4\ep} \,. 
\end{align}
Note that $\zg=1$ at this level due to no one-loop vertex correction to $\gc$. 
It is now straightforward to obtain the beta functions
using Eqs. (\ref{eq:zff}-\ref{eq:zvf}), 
\begin{align}
\label{eq:beta_g}
\betg &= - \rb \gcr + \frac{3}{2} \gcr\up{3} + \frac{3}{8} \gcr \gamr\up{2} + \frac{1}{2} \gvr\up{2} \gcr \,, \\
\label{eq:beta_gam}
\betgam &= - \frac{\ep}{2} \gamr + \gamr\up{3} + \gcr\up{2} \gamr \,, \\
\label{eq:beta_v}
\betv &= - \frac{\epp}{2} \gvr + \gcr\up{2} \gvr \,.
\end{align}


\subsection{Fixed points and stability}
\label{sec:fp}

By analyzing where the beta functions vanish, we obtain the following fixed points, 
(FP $\equiv (\gcr*\up{2}, \gamr*\up{2}, \gvr*\up{2})$):
\begin{align}
\label{eq:fp1}
FP_{1} &= (0,0,0) \,, \\
\label{eq:fp2}
FP_{2} &= \left(0, \frac{\ep}{2}, 0 \right) \,, \\
\label{eq:fp3}
FP_{3} &= \left( \frac{2\rb}{3}, 0, 0 \right)  \,, \\
\label{eq:fp4}
FP_{4} &= \left( \frac{\epp}{2}, 0, 2\rb - \frac{3}{2} \epp \right)  \,, \\
\label{eq:fp5}
FP_{5} &= \left( -\frac{\ep}{6}+\frac{8\rb}{9}, \frac{2\ep}{3}-\frac{8\rb}{9}, 0 \right) \,, \\
\label{eq:fp6}
FP_{6} &= \left( \frac{\epp}{2}, \frac{\ep}{2}-\frac{\epp}{2}, 2\rb - \frac{3}{8}\ep - \frac{9}{8}\epp \right)  \,.
\end{align}
{\change Apart from the Gaussian fixed point, $FP_1$, we find five other fixed points. The fixed points $FP_2$ and $FP_3$ have been studied earlier in the context of an impurity spin \cite{SBV1999,VBS2000,SS2001} and Kondo-impurity Hamiltonian \cite{VojtaFritz04,FritzVojta04} respectively. The fixed point $FP_5$ is the deconfined critical point found in Ref. \cite{dqcp_tj}. Here we find two additional fixed points, $FP_4$ and $FP_6$. }
For $FP_{5}$ to be real, we need $3\ep/8 < 2\rb < 3\ep/2$. While for $FP_{6}$ to be real we need $\ep>\epp>0$ and $2\rb > (3\ep + 9\epp)/8$. {\change Similarly, the reality condition for other fixed points is straightforward to see.}

We will now do the stability analysis of the fixed points by looking at the eigenvalues of the following stability matrix:
\begin{equation}
\label{eq_J_mat}
J \equiv 
\begin{bmatrix}
J_{1} & J_{2} & J_{3} \\
J_{4} & J_{5} & J_{6} \\
J_{7} & J_{8} & J_{9} 
\end{bmatrix} \,,
\end{equation}
where,
\begin{align}
\label{eq:Jdef}
J_{1} &\equiv \frac{\partial \betg}{\partial \gcr} = -\rb + \frac{9}{2} \gcr\up{2} + \frac{3}{8} \gamr\up{2} 
+ \frac{\gvr\up{2}}{2} \,, ~~~~
J_{2} \equiv \frac{\partial \betg}{\partial \gamr} = \frac{3}{4} \gcr \gamr \,, ~~~~ 
J_{3} \equiv \frac{\partial \betg}{\partial \gvr} = \gvr\gcr \,, \nonumber \\ 
J_{4} &\equiv \frac{\partial \betgam}{\partial \gcr} = 2\gcr \gamr \,, ~~~~
J_{5} \equiv \frac{\partial \betgam}{\partial \gamr} = -\frac{\ep}{2} + 3\gamr\up{2} + \gcr\up{2} \,, ~~~~
J_{6} \equiv \frac{\partial \betgam}{\partial \gvr} = 0 \,, \nonumber \\
J_{7} &\equiv \frac{\partial \betv}{\partial \gcr} = 2\gcr \gvr \,, ~~~~
J_{8} \equiv \frac{\partial \betv}{\partial \gamr} = 0 \,, ~~~~
J_{9} \equiv \frac{\partial \betv}{\partial \gvr} = -\frac{\epp}{2} + \gcr\up{2} \,.
\end{align}
From the eigenvalues of the above matrix {\change (see Appendix \ref{app:eig})}, it is immediately clear that for $\rb>0$, $\ep>0$ and $\epp>0$, the Gaussian fixed point $FP_{1}$ is always unstable. 

For $FP_{5}$ to be a stable fixed point, we require $\ep>0$, $3\ep/8 < 2\rb < 3\ep/2$, and $2\rb > (3\ep + 9\epp)/8$. The second inequality is trivially satisfied as soon as $FP_{5}$ is real. If we use in addition the self-consistency condition $\ep=2\rb=1$ (to be discussed in Section~\ref{sec:eta_scn}), this implies that $FP_{5}$ is stable if $\epp<5/9$ (although we cannot trust the present expansion at values of $\epp$ of order unity). 

For $FP_{6}$ the eigenvalues of the stability matrix are given by the following characteristic polynomial: $\lambda\up{3} + A\lambda\up{2} + B\lambda + C$. The corresponding coefficients are as follows:
\begin{equation}
A = -\ep - \frac{\epp}{2} \,,~~~ B = \epp (\frac{3\ep}{2} - 2\rb) \,,~~~ 
C = \frac{\epp}{8} (\ep-\epp) (16\rb - 3\ep - 9\epp) \,.
\end{equation}
From the condition for $FP_{6}$ to be real it is clear that $C>0$ which implies that at least one eigenvalue is negative if $FP_{6}$ is real. Therefore the non-trivial fixed point $FP_{6}$ is unstable. If this fixed point is real it always has one relevant direction. 
{\change We also note that the other new fixed point, $FP_{4}$, found in this work also has at least one unstable direction as soon as it is real. }


\subsection{Anomalous dimension of $f$ and $b$ operators}
\label{sec:eta_fb}

We now calculate the anomalous dimension of the $f$ and $b$ propagators, defined as follows:
\begin{equation}
\label{eq:adim}
\eta_{f} = \rgs \frac{d \ln \zf}{d \rgs}|_{FP} \,, ~~~~~ \eta_{b} = \rgs \frac{d \ln \zb}{d \rgs}|_{FP}
\end{equation}
In our case,
\begin{equation}
\label{eq:eta}
\rgs \frac{d \ln \zf}{d \rgs} = \gcr\up{2} + \frac{3}{4} \gamr\up{2} + \gvr\up{2} \,, ~~~~~
\rgs \frac{d \ln \zb}{d \rgs} = 2 \gcr\up{2}  \,.
\end{equation}
Thus we find the following anomalous dimension at the fixed points,
\begin{align}
\label{eq:adim1}
&FP_1: \eta_{f} = 0 \,,~~~~ \eta_{b} = 0 \,, \\
\label{eq:adim2}
&FP_2: \eta_{f} = \frac{3}{8}\ep \,,~~~~ \eta_{b} = 0 \,, \\
\label{eq:adim3}
&FP_3: \eta_{f} = \frac{2}{3}\rb \,,~~~~ \eta_{b} = \frac{4}{3}\rb \,, \\
\label{eq:adim4}
&FP_4: \eta_{f} = 2\rb - \epp \,,~~~~ \eta_{b} = \epp \,, \\
\label{eq:adim5}
&FP_5: \eta_{f} = \frac{1}{3}\ep + \frac{2}{9}\rb \,,~~~~ \eta_{b} = -\frac{1}{3}\ep + \frac{16}{9}\rb \,, \\
\label{eq:adim6}
&FP_6: \eta_{f} = 2\rb - \epp \,,~~~~ \eta_{b} = \epp \,.
\end{align}
{\change However, note that these exponents are not physical observables since the operators $f$ and $b$ are not gauge invariant.}


\subsection{Anomalous dimension of spin, electron and density operators}
\label{sec:eta_scn}

We are interested in the anomalous dimensions of the gauge-invariant operators, $S$, $c$, and $n$. For this purpose we can look at the correlators $\langle\vec{S}(\tau) \cdot \vec{S}(0)\rangle$, 
$\langle c_{\alpha}(\tau) c\up{\dagger}_{\alpha} (0) \rangle$, and $\langle n(\tau) n(0) \rangle$ made from the composite operators $\fd{\al} \sigma\up{a}_{\al \be} \fa{\be}/2$, $\fd{\al} \ba$, and $\fd{\al} \fa{\al}$ respectively. In order to proceed, we first introduce these composite operator terms in the action, such that, \begin{equation}
\label{eq:S_comp}
S(D) = \frac{1}{\beta}  \sum_{i \omega_n} \left( 
\LS \fd{\al} \frac{\sigma\up{a}_{\al \be}}{2} \fa{\be} 
+ \LP [\fd{\al} \ba + H.c.] + \LN \fd{\al} \fa{\al} \right) + S_{rest} (D) \,,
\end{equation}
where $S_{rest}$ has all the other terms in the action analyzed before. 
{\change As we shall see in the following, this procedure will directly yield us the renormalization factors for the required gauge-invariant operators, and consequently their anomalous dimensions.}

We define the renormalized couplings and the renormalized composite operators $\hat{S} = \fd{\al} \frac{\sigma\up{a}_{\al \be}}{2} \fa{\beta}$, 
$c_{\alpha}\up{\dagger} = \fd{\al} \ba$, and $n = \fd{\al} \fa{\al}$ as follows
\begin{align}
\label{eq:lam_renorm2}
\LS &= \frac{Z_{ff} \Lambda_{S,R}}{\zf} \,,~~~~~ \LP = \frac{Z_{fb} \Lambda_{c,R}}{\sqrt{\zf \zb}} \,, ~~~~~ 
\LN = \frac{Z_{ff1} \Lambda_{n,R}}{\zf} \,, \\
\label{eq:sc_renorm}
\hat{S} &= \sqrt{Z_{S}} \hat{S}_{R} \,,~~~~~ c = \sqrt{Z_{c}} c_{R} \,,~~~~~ n = \sqrt{Z_{n}} n_{R} \,.
\end{align}
We find that the diagrams required to evaluate the vertex corrections to $\LS$, $\LP$, and $\LN$ are exactly those that we used in the calculation of $\zgam$, $\zg$, and $\zv$ respectively. Therefore,
\begin{equation}
\label{eq:zs_zc}
Z_{S} = \left( \frac{\zf}{\zgam} \right)\up{2} \,,~~~~~ Z_{c} = \frac{\zf \zb}{\zg\up{2}} \,,~~~~~ 
Z_{n} = \left( \frac{\zf}{\zv} \right)\up{2} \,.
\end{equation}
This readily gives us,
\begin{align}
\label{eq:zs}
Z_{S} &= 1 - \frac{\gcr\up{2}}{\rb} - \frac{2 \gamr\up{2}}{\ep} \,, \\
\label{eq:zc}
Z_{c} &= 1 - \frac{3\gcr\up{2}}{2\rb} - \frac{3 \gamr\up{2}}{4 \ep} - \frac{\gvr\up{2}}{\epp} \,, \\
\label{eq:zn}
Z_{n} &= 1 - \frac{\gcr\up{2}}{\rb} \,.
\end{align}
We can now evaluate the anomalous dimensions as,
\begin{align}
\label{eq:etas_f}
\eta_{S} &\equiv \frac{d\ln Z_{S}}{d\ln \mu} = \frac{1}{Z_{S}} \left[ \frac{\partial Z_{S}}{\partial \gcr} \betg 
+ \frac{\partial Z_{S}}{\partial \gamr} \betgam + \frac{\partial Z_{S}}{\partial \gvr} \betv \right] 
= 2(\gcr\up{2} + \gamr\up{2}) \,, \\
\label{eq:etac_f}
\eta_{c} &\equiv \frac{d\ln Z_{c}}{d\ln \mu} = \frac{1}{Z_{c}} \left[ \frac{\partial Z_{c}}{\partial \gcr} \betg 
+ \frac{\partial Z_{c}}{\partial \gamr} \betgam + \frac{\partial Z_{c}}{\partial \gvr} \betv \right] 
= 3\gcr\up{2} + \frac{3}{4}\gamr\up{2} + \gvr\up{2} \,, \\
\label{eq:etan_f}
\eta_{n} &\equiv \frac{d\ln Z_{n}}{d\ln \mu} = \frac{1}{Z_{n}} \left[ \frac{\partial Z_{n}}{\partial \gcr} \betg 
+ \frac{\partial Z_{n}}{\partial \gamr} \betgam + \frac{\partial Z_{n}}{\partial \gvr} \betv \right] 
= 2\gcr\up{2} \,.
\end{align}
The anomalous dimensions at the fixed points are listed in Table \ref{tab:adim_f}. 
Just as shown in Ref. \cite{dqcp_tj}, we can also make an exact statement here. 
To all orders in $\ep$, $\epp$, and $\rb$: If $\gcr*\neq 0$ then $\eta_{c}=2\rb$, 
if $\gamr*\neq 0$ then $\eta_{S}=\ep$, and if $\gvr*\neq 0$ then $\eta_{n}=\epp$. 
{\change This statement can be easily proved by differentiating the relations for the coupling constants in Eq. (\ref{eq:renorm_factf}) with respect to the RG scale $\mu$ and using the definitions in Eqs. (\ref{eq:etas_f})-(\ref{eq:etan_f}). }
Thus at the non-trivial fixed point, $FP_{6}$, 
$\eta_{S}=\ep$, $\eta_{c}=2\rb$, and $\eta_{n}=\epp$ to all orders in $\ep$, $\epp$ and $\rb$. While at the non-trivial fixed point $FP_{5}$, $\eta_{S}=\ep$ and $\eta_{c}=2\rb$ to all orders, but $\eta_{n}$ can not be evaluated exactly to all orders. 

\begin{table}[t]
\begin{center}
\begin{tabular}{|c||c|c|c|c|c|}
\hline
Fixed point & $\eta_{S}$ & $\eta_{c}$ & $\eta_{n}$ \\ 
\hline
\hline
$FP_{1}$ & $0$ & $0$ & $0$ \\[5pt]  \hline 
$FP_{2}$ & $\ep$ & $\frac{3}{8}\ep$ & $0$ \\[5pt] \hline
$FP_{3}$ & $\frac{4}{3}\rb$ & $2\rb$ & $\frac{4}{3}\rb$ \\[5pt] \hline
$FP_{4}$ & $\epp$ & $2\rb$ & $\epp$ \\[5pt] \hline
$FP_{5}$ & $\ep$ & $2\rb$ & $\frac{16}{9}\rb - \frac{\ep}{3}$ \\[5pt] \hline
$FP_{6}$ & $\ep$ & $2\rb$ & $\epp$ \\[5pt] 
\hline
\end{tabular} 
\caption{Anomalous dimensions at fixed points. }
\label{tab:adim_f}
\end{center}
\end{table}

{\change We now recall the self-consistency condition, Eq. (\ref{eq:self_cons}), which we shall shortly impose at the non-trivial fixed point. Recally that we started out with our RG assuming the power-law behavior for the fields $P$, $Q$, and $R$ (see Eq. (\ref{QRpower})). In the last paragraph we calculated the exponents corresponding to the correlators $\langle\vec{S}(\tau) \cdot \vec{S}(0)\rangle$, $\langle c_{\alpha}(\tau) c\up{\dagger}_{\alpha} (0) \rangle$, and $\langle n(\tau) n(0) \rangle$, which enter the RHS of self-consistency conditions in Eq. (\ref{eq:self_cons}). In order to satisfy the self-consistency conditions in Eq. (\ref{eq:self_cons}) the exponents on the LHS and RHS of the expressions must be the same. Therefore satisfying the self-consistency for $Q$, $R$, and $P$ fields means $\eta_S = 2-\ep$, $\eta_c=2-2\rb$, and $\eta_n=2-\epp$ respectively, where $\eta$s are given by the expressions in Eqs. (\ref{eq:etas_f})-(\ref{eq:etan_f}) or Table \ref{tab:adim_f} (at fixed points). }


At the fixed point $FP_{5}$ (i.e., the DQCP FP from Ref. \cite{dqcp_tj}), 
{\change we impose the self-consistency conditions on $Q$ and $R$, Eq. (\ref{eq:self_cons}), but there is no self-consistency condition on $P$ since $K=0$. Using the above prescription this fixes the values of $\ep=1$  
and $\rb=1/2$ by matching the exponents of $Q$ and $R$ in Eq. (\ref{eq:self_cons}) to those of $\eta_{S}$ and $\eta_{c}$ respectively, found above (see Table \ref{tab:adim_f}). 
However, since there is no self-consistency condition involving $\eta_{n}$ the value of $\epp$ is not fixed. }
Since the exponents $\eta_{c}$ and $\eta_{S}$ are obtained exactly, their values of $\eta_{c}=2\rb=1$ and $\eta_{S}=\ep=1$ can be trusted. But the exponent $\eta_{n}$ is not exact and will have corrections from higher order expansion in $\rb$ and $\ep$ (it does not depend upon $\epp$ at $FP_5$).
We can choose any $\epp<5/9$ so that $FP_{5}$ is stable. We then obtain our main result that $\eta_{n} = 5/9$, using Eq. (\ref{eq:etan_f}) or Table \ref{tab:adim_f} {\change and the self-consistent values of $\ep=2\rb=1$}.

Note that at the other non-trivial fixed point, $FP_{6}$, the exponents $\eta_{c}=2\rb$, $\eta_{S}=\ep$ and $\eta_{n}=\epp$ are obtained exactly. {\change Here we need to impose the self-consistency conditions on all the three fields $P$, $Q$, and $R$. Again following the above prescription, we obtain the self-consistent values of $2\rb=\ep=\epp=1$.}
Hence, at this fixed point $\eta_{c}=\eta_{S}=\eta_{n}=1$. For these large values of $\rb$, $\ep$ and $\epp$ the fixed point $FP_{6}$ becomes complex and is {\change unstable} at one loop order, but there is no justification for using the one loop results at these large values.

{\change Similarly, at the other new fixed point, $FP_4$, the self-consistency conditions yields the values $2\rb = \epp=1$. Here the value of $\ep$ is not fixed. However, for these values this fixed point is complex and unstable at one-loop order.}


\subsection{Flow of $s$}
\label{sec:s}

At one-loop level, we can derive the flow of $s$, {\change which was set to zero at the critical point in the above discussion}.
{\change The parameter $s$ is nothing but the difference between the masses of the $f$ and $b$ fields. Using the standard momentum-shell RG procedure, and the self-energies of $f$ and $b$ fields, it is straight forward to obtain the renormalization of $s$. We refer the interested readers to Appendix (D.1) in Ref. \cite{dqcp_tj} where the technical steps (for $K=0$) are sketched in detail. Following these steps we obtain the beta function of $s$ as follows:}
\begin{equation}
\bets = -s + 3 s \gcr\up{2} - \gcr\up{2} + \frac{3}{4} \gamr\up{2} + \gvr\up{2} \,.
\end{equation}
This governs the flow away from the critical point, discussed above for $s_0 = 0$. It turns out that $s$ is always a relevant parameter. {\change As shown in Ref. \cite{dqcp_tj}, $s$ tunes the phase transition from a metallic spin glass phase to a disordered Fermi liquid \cite{dqcp_tj} }. 


\section{Conclusion}
\label{sec:conc}

This paper has presented a renormalization group analysis of the $t$-$J$-$K$ model in (\ref{eq:ham_tJK}), a model for the cuprates with random and infinite-range interactions. This model was previously studied without the density-density interaction, $K$, in Ref.~\onlinecite{dqcp_tj}: they found a deconfined critical point at a non-zero doping $p=p_c$, separating a metallic spin glass for $p<p_c$, from a disordered Fermi liquid for $p> p_c$. In the present paper, we examined the fate of this fixed point for non-zero $K$, and also computed the exponent characterizing density correlations. {\change To our knowledge, a microscopic calculation of this quantity has not been done before, and our calculations are relevant to cuprates and related materials.}

Recent momentum-resolved electron energy-loss spectroscopy (M-EELS) experiments \cite{Mitrano18,Husain19} have observed anomalous density fluctuations near optimal doping in the cuprates. In our theory, the critical density fluctuations are characterized by the  spectral density
\beq
\chi_n^{\prime\prime} (\omega) \sim \mbox{sgn}(\omega) |\omega|^{\eta_n - 1} \quad, \quad T=0, \label{chin}
\eeq
and similarly for the spin fluctuations with exponent $\eta_S$. 
{\change These spectral functions are obtained from the imaginary part of the respective correlation functions.
At non-zero $T$, the spectrum is characterized by a `Planckian' frequency scale, and (\ref{chin}) is multiplied by a universal function of $\hbar \omega/(k_B T)$ so that we can write
\beq
\chi_n^{\prime\prime} (\omega) \sim T^{\eta_n - 1} \Phi_{\eta_n}
\left( \frac{\hbar \omega}{k_B T} \right) \,; 
\label{chin2}
\eeq
(\ref{chin}) holds for $\hbar \omega \gg k_B T$, while $\chi_n^{\prime\prime} \sim \omega/T^{2-\eta_n}$ for $\hbar \omega \ll k_B T$. 
The explicit form of the function $\Phi_{\eta}$ can be determined by conformal mapping \cite{SSS94,PGKS97,PG98}
\beq
\Phi_\eta (y) = \sinh \left( \frac{y}{2} \right) \left| \Gamma \left( \frac{\eta}{2} + \frac{i y}{2 \pi} \right) \right|^2 \,.
\label{chin3}
\eeq
We note that in a Fermi liquid $\Phi_2(y) = y/2$ is a linear function, so that $\chi_n^{\prime\prime} (\omega) \sim \omega$ is $T$-independent. All other value of $\eta_n$ yield a non-trivial $T$ dependence, including the marginal case, for which $\Phi_1 (y) = \pi \tanh(y/2)$.}

{\change The M-EELS experiments \cite{Mitrano18,Husain19} seem to observe a frequency independent density response at the optimal doping. In terms of the spectral density (\ref{chin}), this corresponds to having the exponent $\eta_n =1$. }
In this paper, we found a new fixed point, $FP_6$, with $K \neq 0$, at which the exponents can be determined to all loop order: we obtained the `marginal' value $\eta_n = \eta_S = 1$. However, at least the one loop order at which our computations were carried out, this fixed point was unstable to the previously found \cite{dqcp_tj} fixed point at $K=0$, labeled $FP_5$ here. But it cannot be ruled out that at strong coupling $FP_6$ is the appropriate fixed point, and we expect  $\eta_n = \eta_S = 1$ to continue to hold exactly at any such fixed point with $K \neq 0$. 
{\change Therefore our theory provides a possible route to explain the origin of the exponent $\eta_n=1$ observed in the experiments. }

At the $K=0$ fixed point $FP_5$, we previously showed that $\eta_S = 1$ to all loop order \cite{dqcp_tj}.
In the present paper, we are only able to determine $\eta_n$ at $FP_5$ to one loop (there is no corresponding argument to extend the computation of $\eta_n$ to all orders): the result is shown in 
Table~\ref{tab:adim_f}. 
At the self-consistent values of the expansion parameters, $\epsilon=2\rb=1$, 
the exponent evaluates to $\eta_n = 5/9$. 
{\change 
However, our computation is first order in $\epsilon$, $\rb$ (both of the same order), and so we expect corrections to the value quoted here. } 

We hope that numerical studies of Hamiltonians like (\ref{eq:ham_tJK}) will shed further light on the existence and nature of the finite doping deconfined critical point.


\section*{Acknowledgement}
\label{sec:ack}

We thank Chenyuan Li, Grigory Tarnopolsky, and Antoine Georges for collaboration on previous work \cite{dqcp_tj}. We also thank P.~Abbamonte and M.~Mitrano for discussions of the M-EELS experiments.
This research was supported by the National Science Foundation under Grant No.~DMR-2002850. 
D.G.J acknowledges support from the Leopoldina fellowship by the German National Academy of Sciences through grant no. LPDS 2020-01. This work was also supported by the Simons Collaboration on Ultra-Quantum Matter, which is a grant from the Simons Foundation (651440, S.S.).


\appendix 

{\change
\section{Eigenvalues of stability matrix}
\label{app:eig}

Here we quote the eigenvalues of the stability matrix (\ref{eq_J_mat}) evaluated at the fixed points. 
\begingroup
\allowdisplaybreaks
\begin{align}
&FP_1 : \left\lbrace -\rb, -\frac{\ep}{2}, -\frac{\epp}{2} \right\rbrace \,, \\
&FP_2 : \left\lbrace \ep, \frac{3 \ep - 16 \rb}{16}, -\frac{\epp}{2} \right\rbrace \,, \\
&FP_3 : \left\lbrace 2\rb, \frac{4\rb - 3\ep}{6}, \frac{4\rb - 3\ep}{6} \right\rbrace \,, \\
&FP_4 : \left\lbrace \frac{\epp - \ep}{2}, \frac{1}{4} \left(3\epp - \sqrt{32\rb\epp - 15 \ep\up{'2}} \right), 
\frac{1}{4} \left(3\epp + \sqrt{32\rb\epp - 15 \ep\up{'2}} \right) \right\rbrace \,, \\
&FP_5 : \left\lbrace \frac{1}{36} \left( 16 \rb + 15 \ep - \sqrt{4864 \rb\up{2} - 3840 \rb \ep + 873 \ep\up{2} } \right), 
\frac{1}{36} \left( 16 \rb + 15 \ep + \sqrt{4864 \rb\up{2} - 3840 \rb \ep + 873 \ep\up{2} } \right), \right. \nonumber \\
&~~~~~~~~~~~~~~~~~~~~~~~~~~~ \left. \frac{1}{18} \left( 16 \rb - 3 \ep - 9 \epp \right) \right\rbrace \,.
\end{align}
\endgroup
The eigenvalues at $FP_6$ are discussed in the main text using its characteristic polynomial.
}

\section{RG in terms of gauge-invariant operators}
\label{sec:rg_scn}

In this appendix we present an alternative RG analysis directly in terms of the gauge-invariant operators. This also has the advantage that we can present our results for a general $M$ and $M'$, which generalizes SU$(1|2)$ to SU$(M'|M)$.
We have the following impurity {\change and bath} Hamiltonian as before,
\begin{align}
H_{\rm imp} &= \gc \left( c\up{\dagger}_{\ell \alpha} \psi_{\alpha \ell} (0) + H.c. \right) 
+ \gam S\up{a} \phi_{a} (0) + \gv \nt \zeta(0)  \nonumber \\
&+ \int |k|\up{r} dk k \psi\up{\dagger}_{k \alpha \ell} \psi_{k \alpha \ell} 
+ \frac{1}{2} \int d\up{d}x \left[ \pi_{a}\up{2} + (\partial_{x} \phi_{a})\up{2} \right] 
+ \frac{1}{2} \int d\up{d'}x \left[ \tilde{\pi}\up{2} + (\partial_{x} \zeta)\up{2} \right] \,,
\end{align}
where $\alpha = 1, ..., M$, $\ell = 1, ..., M'$ and $a = 1, ..., M\up{2}-1$. This Hamiltonian is a large $M$, $M'$ generalization of Eq. \ref{eq:H_imp}. In the above Hamiltonian, $\nt \equiv n - \nf$ with $n\equiv \fd{\alpha} \fa{\alpha}$ and $\nf \equiv \langle \fd{\alpha} \fa{\alpha} \rangle_{0} = 2/3$. 
To proceed with RG, we first introduce the following renormalization factors,
\begin{align}
\label{eq:renorm_fact_m}
S\up{a}&=\sqrt{Z_{S}} S\up{a}_{R}\,, ~~~ c_{p\alpha}=\sqrt{Z_{c}}c_{R,p\alpha} \,, ~~~ \nt = \sqrt{Z_{\nt}} \nt_{R} \,, ~~~ 
n = \sqrt{Z_{n}} n_{R} \,, \nonumber \\
\gam &= \frac{\mu\up{\ep/2} \tilde{Z}_{\gamma}}{\sqrt{Z_{S} \tilde{S}_{d+1}}} \gamr \,, ~~~
\gc= \frac{\mu\up{\rb} \tilde{Z}_{g}}{\sqrt{Z_{c} \Gamma(r+1)}} \gcr \,, ~~~
\gv= \frac{\mu\up{\epp/2} \tilde{Z}_{v}}{\sqrt{Z_{\nt} \tilde{S}_{d'+1}}} \gvr \,.
\end{align}
In what follows we will also make use of the following expression for expectation values:
\begin{align}
\mathcal{I}_{m,m'}  &\equiv \left\langle \left(f_\alpha^\dagger f_\alpha\right)^m \left( b_\ell^\dagger b_\ell \right)^{m'} \right\rangle  \nonumber \\
&= \frac{1}{\mathcal{D}(M, M',P)} \oint_{|z|=c < 1} \frac{dz}{2 \pi i} \frac{1}{z^{P+1}} \,
\left[ \left( z \frac{d}{dz} \right)^{m} (1+z)^M \right] \left[ \left( z \frac{d}{dz} \right)^{m'} \frac{1}{(1-z)^{M'}} \right]  \,.
\end{align}
For more details we refer to Ref. \cite{dqcp_tj}. We just recall that $\mathcal{I}_{0,0} = 1$ and the values for $M=2$, $P=1$, and $M'=1$, which is the case of interest to us are as follows:
\bea
\mathcal{I}_{m,0} &=& \frac{2}{3}, \quad m \geq 1; \quad
\mathcal{I}_{0,m'}  = \frac{1}{3}, \quad m' \geq 1; \quad
\mathcal{I}_{m,m'} = 0, \quad m \geq 1 ~{\rm and}~ m' \geq 1 \,.
\eea

\subsection{Spin correlator}
\label{sec:ss}

\begin{figure}[t]
\centering
\includegraphics[width=0.8\textwidth]{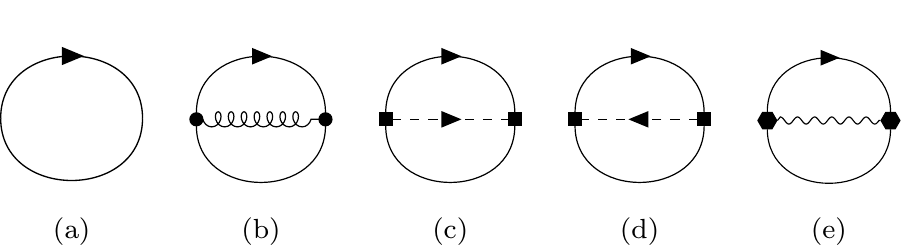}
\caption{Diagrams used to evaluate the denominator, $D$ (Eq. \ref{eq:d}). Note that these are not Feynman diagrams (see the text and Ref. \cite{dqcp_tj} for details). Here the solid line denotes the imaginary time trajectory of the SU($M'|M$) superspin. A filled circle represents a $\gam$ vertex, a filled square represents a $\gc$ vertex, and a filled hexagon represents a $\gv$ vertex. The $\phi$, $\psi$, and $\zeta$ propagators are represented by a spiral curve, a dashed curve, and a wiggly curve respectively.}
\label{fig:denom}
\end{figure}

\begin{figure}[t]
\centering
\includegraphics[width=0.95\textwidth]{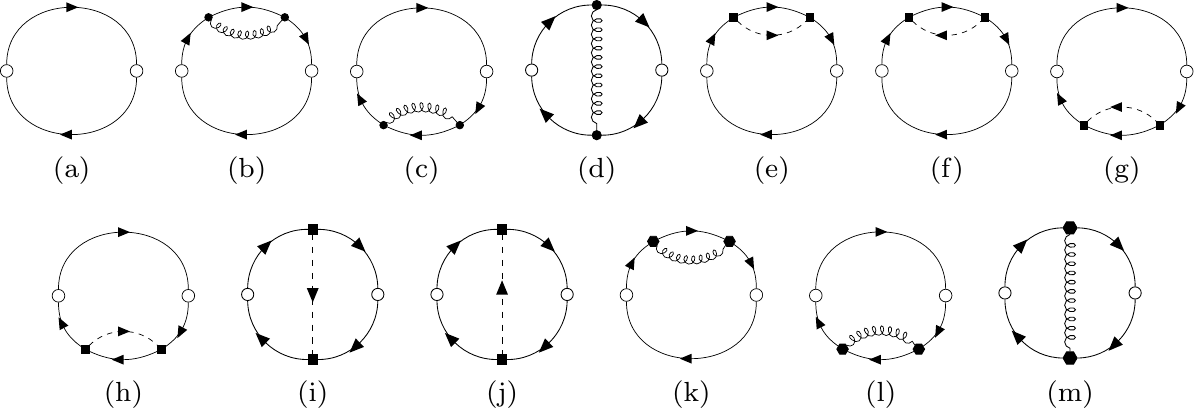}
\caption{Diagrams used in the evaluation of the numerator, $N_{1}$ (Eq. \ref{eq:n1}), of $\langle O_{1} \rangle = \langle S\up{a}(\tau) S\up{a}(0) \rangle$. Here, the external $S\up{a}$ operator is represented by an open circle. Apart from this the rest of the conventions are same as in Fig.~\ref{fig:denom}.} 
\label{fig:N1}
\end{figure}

Here we calculate the spin correlator, $\langle O_{1} \rangle \equiv \langle S\up{a}(\tau) S\up{a}(0) \rangle$, which will give us $Z_{S}$. We will follow the strategy from Ref. \cite{VBS2000, dqcp_tj}, which relies on explicit evaluation of operator traces rather than the Wick's theorem, such that $\langle O_{1} \rangle = N_{1}/D$. We evaluate the denominator and numerator in $\langle O_{1} \rangle$ using the diagrams shown in Figs. \ref{fig:denom} and \ref{fig:N1} respectively to obtain,
\begin{align}
\label{eq:d}
D &= 1 + \gam\up{2} \lo \left( \Da + \Db + \Dc \right) + \gc\up{2} \lop \left( \Dap + \Dbp + \Dcp \right) \nonumber \\
&~~~~~~~~~~+ \gc\up{2} \lopp \left( \Dapp + \Dbpp + \Dcpp \right) 
+ \gv\up{2} \loppp \left( \Daz + \Dbz + \Dcz \right) \,, \\
\label{eq:n1}
N_{1} &= \lo + \gam\up{2} \left( \la\Da + \lb\Db + \lc\Dc \right) 
+ \gc\up{2} \left( \lap\Dap + \lbp\Dbp + \lcp\Dcp \right) \nonumber \\
&~~~~~~~~+ \gc\up{2} \left( \lapp\Dapp + \lbpp\Dbpp + \lcpp\Dcpp \right) 
+ \gv\up{2} \left( \lappp\Daz + \lbppp\Dbz + \lcppp\Dcz \right) \,.
\end{align}

The diagrams in Figs. \ref{fig:denom} (a)-(d) and Figs. \ref{fig:N1} (a)-(j) have been evaluated before in Ref. \cite{dqcp_tj}. The expressions for $L_{i}$, $L'_{i}$ and $L''_{i}$ can be found in Eqs. (B5)-(B16) in Ref. \cite{dqcp_tj}, while those for $D_{i}$, $D'_{i}$ and $D''_{i}$ can be found in Eqs. (B17)-(B25) in Ref. \cite{dqcp_tj}. We quote here the previously not evaluated expressions,
\begingroup
\allowdisplaybreaks
\begin{align}
\loppp &= \lbr \nt \nt \rbr = \mathcal{I}_{2,0} - 2\nf \mathcal{I}_{1,0} + \nf\up{2} \,, \\
\lappp &= \lbr S\up{a} \nt \nt S\up{a} \rbr = \frac{M+1}{2M}(M \mathcal{I}_{3,0} - \mathcal{I}_{4,0} - 2\nf(M\mathcal{I}_{2,0} - 
\mathcal{I}_{3,0}) + \nf\up{2}(M\mathcal{I}_{1,0} - \mathcal{I}_{2,0}) ) \,, \\
\lbppp &= \lbr S\up{a} S\up{a} \nt \nt \rbr = \frac{M+1}{2M}(M \mathcal{I}_{3,0} - \mathcal{I}_{4,0} - 2\nf(M\mathcal{I}_{2,0} - 
\mathcal{I}_{3,0}) + \nf\up{2}(M\mathcal{I}_{1,0} - \mathcal{I}_{2,0}) ) \,, \\
\lcppp &= \lbr S\up{a} \nt S\up{a} \nt \rbr = \frac{M+1}{2M}(M \mathcal{I}_{3,0} - \mathcal{I}_{4,0} - 2\nf(M\mathcal{I}_{2,0} - 
\mathcal{I}_{3,0}) + \nf\up{2}(M\mathcal{I}_{1,0} - \mathcal{I}_{2,0}) ) \,. 
\end{align}
\endgroup

Also,
\begingroup
\allowdisplaybreaks
\begin{align}
%
\Daz &= \int_{0}\up{\tau} d\tau_{1} \int_{\tau_{1}}\up{\tau} d\tau_{2} G_{\zeta} (\tau_{1} - \tau_{2})  
= - \frac{\widetilde{S}_{d'+1} \tau\up{\epp}}{\epp (1-\epp)} \,, \\
\Dbz &= \int_{\tau}\up{\beta} d\tau_{1} \int_{\tau_{1}}\up{\beta} d\tau_{2} G_{\zeta} (\tau_{1} - \tau_{2}) 
= - \frac{\widetilde{S}_{d'+1} \tau\up{\epp}}{\epp (1-\epp)} \,, \\
\Dcz &= \int_{0}\up{\tau} d\tau_{1} \int_{\tau}\up{\beta} d\tau_{2} G_{\zeta} (\tau_{1} - \tau_{2})  
= \frac{2 \widetilde{S}_{d'+1} \tau\up{\epp}}{\epp (1-\epp)} \,, \\
G_{\zeta}(\tau) &= \int \frac{d\up{d'}k}{(2\pi)\up{d'}} \frac{d\omega}{2\pi} \frac{e\up{-i\omega \tau}}{k\up{2} + \omega\up{2}} 
= \frac{\widetilde{S}_{d'+1}}{|\tau|\up{d'-1}} \,.
\end{align}
\endgroup 

Using Eqs. \ref{eq:d} and \ref{eq:n1} we get,
\begin{align}
\langle O_{1} \rangle = \frac{N_{1}}{D} &= \lo \bigg \lbrace 
1 + \gam\up{2} \left[ \left( \frac{\la}{\lo} - \lo \right) \Da 
+ \left( \frac{\lb}{\lo} - \lo \right) \Db 
+ \left( \frac{\lc}{\lo} - \lo \right) \Dc \right]  \nonumber \\
&+ \gc\up{2} \left[ \left( \frac{\lap}{\lo} - \lop \right) \Dap 
+ \left( \frac{\lbp}{\lo} - \lop \right) \Dbp 
+ \left( \frac{\lcp}{\lo} - \lop \right) \Dcp \right] \nonumber \\
&+ \gc\up{2} \left[ \left( \frac{\lapp}{\lo} - \lopp \right) \Dapp 
+ \left( \frac{\lbpp}{\lo} - \lopp \right) \Dbpp 
+ \left( \frac{\lcpp}{\lo} - \lopp \right) \Dcpp \right] \nonumber \\
&+ \gv\up{2} \left[ \left( \frac{\lappp}{\lo} - \loppp \right) \Daz 
+ \left( \frac{\lbppp}{\lo} - \loppp \right) \Dbz 
+ \left( \frac{\lcppp}{\lo} - \loppp \right) \Dcz \right] 
\bigg \rbrace \,.
\end{align}
We thus obtain, 
\begin{equation}
Z_{S} = 1 - \frac{\gamr\up{2}}{\ep} \Lgam - \frac{\gcr\up{2}}{2\rb} \Lg - \frac{\gvr\up{2}}{\epp} \Lv \,,
\end{equation}
where ,
\begin{align}
\Lgam &= \frac{\la + \lb -2\lc}{\lo} \,, \\
\Lg &= \frac{\lap+\lapp+\lbp+\lbpp-2\lcp-2\lcpp}{\lo} \,, \\
\Lv &= \frac{\lappp + \lbppp -2\lcppp}{\lo} \,.
\end{align}
We find that $\Lgam=\Lg=2$ and $\Lv=0$ for $M=2\,, M'=1$. Thus, for $M=2\,, M'=1$,
\begin{equation}
\label{eq:zs_m}
Z_{S} = 1 - \frac{2 \gamr\up{2}}{\ep} - \frac{\gcr\up{2}}{\rb} \,.
\end{equation}


\subsection{Electron correlator}
\label{sec:cc}

\begin{figure}[t]
\centering
\includegraphics[width=0.95\textwidth]{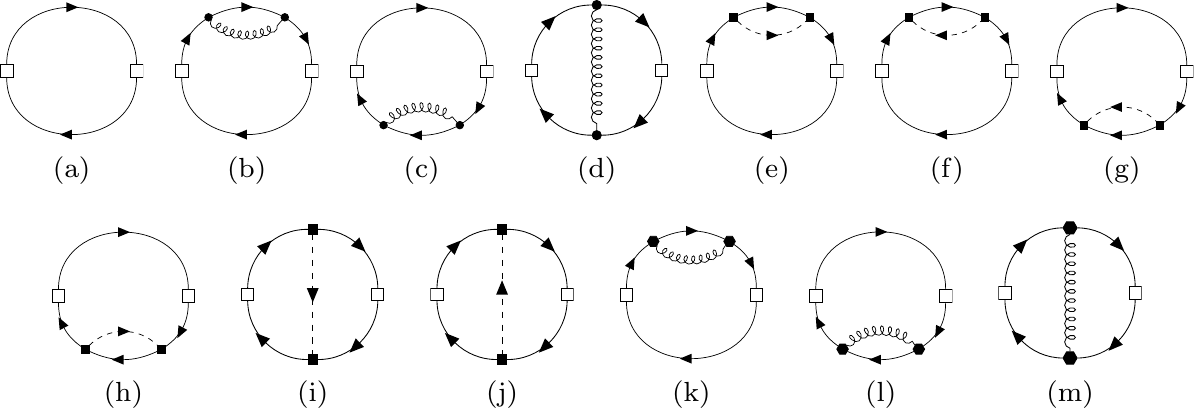}
\caption{Diagrams used in the evaluation of the numerator, $N_{2}$ (Eq. \ref{eq:n2}), of $\langle O_{2} \rangle = \langle c(\tau) c\up{\dagger}(0) \rangle$. Here, the external $c/c\up{\dagger}$ operator is represented by an open square, while the rest of the conventions are the same as in Fig.~\ref{fig:denom}.} 
\label{fig:N2}
\end{figure}

In this subsection we will calculate the electron correlation, $\langle O_{2} \rangle \equiv \langle c(\tau) c\up{\dagger}(0) \rangle = N_{2}/D$. The denominator, $D$, has been already evaluated in Eq. \ref{eq:d}. The numerator, $N_{2}$, is evaluated using the diagrams shown in Fig. \ref{fig:N2}. Thus we obtain,

\begingroup
\allowdisplaybreaks
\begin{align}
\label{eq:n2}
N_{2} &= \po + \gam\up{2} \left( \Pa\Da + \pb\Db + \pc\Dc \right) 
+ \gc\up{2} \left( \pap\Dap + \pbp\Dbp + \pcp\Dcp \right) \nonumber \\
&~~~~~~~+ \gc\up{2} \left( \papp\Dapp + \pbpp\Dbpp + \pcpp\Dcpp \right) 
+ \gv\up{2} \left( \pappp\Daz + \pbppp\Dbz + \pcppp\Dcz \right)\,. 
\end{align}
The diagrams in Fig. \ref{fig:N2} (a)-(j) have been previously evaluated. The expressions for $P_{i}$, $P'_{i}$ and $P''_{i}$ can be found in Eqs. (B33)-(B42) in Ref. \cite{dqcp_tj}. For the rest we have,
\begin{align}
\pappp &= \lbr c\up{\dagger}_{\ell\alpha} \nt \nt c_{\ell\alpha} \rbr =  M' (\mathcal{I}_{3,0}-2\mathcal{I}_{2,0}+\mathcal{I}_{1,0} 
-2\nf(\mathcal{I}_{2,0} - \mathcal{I}_{1,0}) + \nf\up{2}\mathcal{I}_{1,0} ) \nonumber \\ 
&~~~~~~~~~~~~~~~~~+ \mathcal{I}_{3,1}-2\mathcal{I}_{2,1}+\mathcal{I}_{1,1} 
-2\nf(\mathcal{I}_{2,1} - \mathcal{I}_{1,1}) + \nf\up{2}\mathcal{I}_{1,1}   \,, \\
\pbppp &= \lbr c\up{\dagger}_{\ell\alpha} c_{\ell\alpha} \nt \nt \rbr = M' (\mathcal{I}_{3,0} - 2\nf \mathcal{I}_{2,0} + 
\nf\up{2}\mathcal{I}_{1,0}) + \mathcal{I}_{3,1} - 2\nf \mathcal{I}_{2,1} + \nf\up{2}\mathcal{I}_{1,1}  \,, \\
\pcppp &= \lbr c\up{\dagger}_{\ell\alpha} \nt c_{\ell\alpha} \nt \rbr = M' (\mathcal{I}_{3,0}-\mathcal{I}_{2,0} 
-\nf(2 \mathcal{I}_{2,0} - \mathcal{I}_{1,0} ) + \nf\up{2}\mathcal{I}_{1,0} ) \nonumber \\ 
&~~~~~~~~~~~~+ \mathcal{I}_{3,1}-\mathcal{I}_{2,1} - \nf(2 \mathcal{I}_{2,1} - \mathcal{I}_{1,1} ) + \nf\up{2}\mathcal{I}_{1,1} \,.
\end{align}
\endgroup
From Eqs. \ref{eq:d} and \ref{eq:n2} we have,
\begin{align}
\langle O_{2} \rangle = \frac{N_{2}}{D} &= \po \bigg \lbrace 
1 + \gam\up{2} \left[ \left( \frac{\Pa}{\po} - \lo \right) \Da 
+ \left( \frac{\pb}{\po} - \lo \right) \Db 
+\left( \frac{\pc}{\po} - \lo \right) \Dc \right]  \nonumber \\
&+ \gc\up{2} \left[ \left( \frac{\pap}{\po} - \lop \right) \Dap 
+ \left( \frac{\pbp}{\po} - \lop \right) \Dbp 
+\left( \frac{\pcp}{\po} - \lop \right) \Dcp \right] \nonumber \\
&+ \gc\up{2} \left[ \left( \frac{\papp}{\po} - \lopp \right) \Dapp 
+ \left( \frac{\pbpp}{\po} - \lopp \right) \Dbpp 
+\left( \frac{\pcpp}{\po} - \lopp \right) \Dcpp \right] \nonumber \\
&+ \gv\up{2} \left[ \left( \frac{\pappp}{\po} - \loppp \right) \Daz 
+ \left( \frac{\pbppp}{\po} - \loppp \right) \Dbz 
+\left( \frac{\pcppp}{\po} - \loppp \right) \Dcz \right]
\bigg \rbrace \,.
\end{align}
Thus we obtain, 
\begin{equation}
Z_{c} = 1 - \frac{\gamr\up{2}}{\ep} \Pgam - \frac{\gcr\up{2}}{2\rb} \Pg - \frac{\gvr\up{2}}{\ep} \Pv \,,
\end{equation}
where
\begin{align}
\Pgam &= \frac{\Pa + \pb -2\pc}{\po} \,, \\
\Pg &= \frac{\pap + \pbp - 2\pcp + \papp + \pbpp - 2\pcpp}{\po} \,, \\
\Pv &= \frac{\pappp + \pbppp -2\pcppp}{\po} \,.
\end{align}
We obtain $\Pg=3$, $\Pgam=3/4$ and $\Pv=1$ for $M=2\,, M'=1$. Thus, for $M=2\,, M'=1$,
\begin{equation}
\label{eq:zc_m}
Z_{c} = 1 - \frac{3}{4} \frac{\gamr\up{2}}{\ep} - \frac{3}{2} \frac{\gcr\up{2}}{\rb} - \frac{\gvr\up{2}}{\epp} \,.
\end{equation}


\subsection{Density correlator}
\label{sec:nn}

\begin{figure}[t]
\centering
\includegraphics[width=0.95\textwidth]{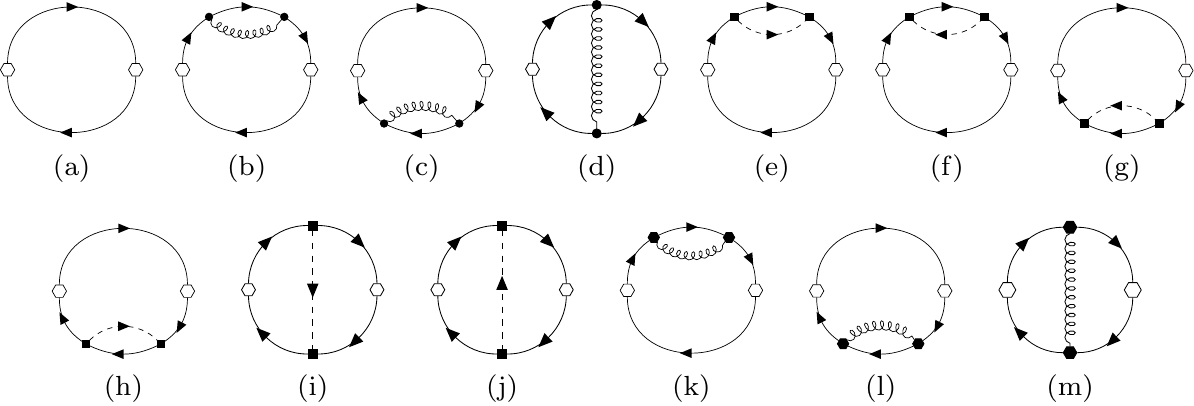}
\caption{Diagrams used in the evaluation of the numerator, $N_{4}$ (Eq. \ref{eq:n4}), of $\langle O_{4} \rangle = \langle n(\tau) n(0) \rangle$. Here, the external $n$ operator is represented by an open hexagon, while the rest of the conventions are same as in Fig.~\ref{fig:denom}.} 
\label{fig:N4}
\end{figure}

In this subsection we will evaluate the density correlation, $\langle O_{4} \rangle \equiv \langle n(\tau) n(0) \rangle = N_{4}/D$. Apart from a constant $\langle \nt(\tau) \nt(0) \rangle$ has the same form as $\langle n(\tau) n(0) \rangle$. The numerator, $N_{4}$, is evaluated using the diagrams shown in Fig. \ref{fig:N4}. We thus have,

\begingroup
\allowdisplaybreaks
\begin{align}
\label{eq:n4}
N_{4} &= \To + \gam\up{2} \left( \Ta\Da + \Tb\Db + \Tc\Dc \right) 
+ \gc\up{2} \left( \Tap\Dap + \Tbp\Dbp + \Tcp\Dcp \right) \nonumber \\
&~~~~~~~+ \gc\up{2} \left( \Tapp\Dapp + \Tbpp\Dbpp + \Tcpp\Dcpp \right) 
+ \gv\up{2} \left( \Tappp\Daz + \Tbppp\Dbz + \Tcppp\Dcz \right)\,, 
\end{align}
where,
\begin{align}
\To &= \lbr nn \rbr = \mathcal{I}_{2,0} \,,\\
\Ta &= \lbr n S\up{a} S\up{a} n \rbr = \frac{M+1}{2M}(M\mathcal{I}_{3,0} - \mathcal{I}_{4,0} )  \,, \\
\Tb &= \lbr n n S\up{a} S\up{a} \rbr = \frac{M+1}{2M}(M\mathcal{I}_{3,0} - \mathcal{I}_{4,0} )  \,, \\
\Tc &= \lbr n S\up{a} n S\up{a} \rbr = \frac{M+1}{2M}(M\mathcal{I}_{3,0} - \mathcal{I}_{4,0} ) \,, \\
\Tap &= \lbr n c_{\ell'\beta} c_{\ell'\beta}\up{\dagger} n \rbr = M \mathcal{I}_{2,1} - \mathcal{I}_{3,1}  \,, \\
\Tbp &= \lbr n n c_{\ell'\beta} c\up{\dagger}_{\ell'\beta} \rbr = M \mathcal{I}_{2,1} - \mathcal{I}_{3,1}  \,, \\
\Tcp &= \lbr n c_{\ell'\beta} n c\up{\dagger}_{\ell'\beta} \rbr = M \mathcal{I}_{1,1} + (M-1) \mathcal{I}_{2,1} - \mathcal{I}_{3,1}  \,, \\
\Tapp &= \lbr n c_{\ell'\beta}\up{\dagger} c_{\ell'\beta} n \rbr = M' \mathcal{I}_{3,0} + \mathcal{I}_{3,1} \,, \\
\Tbpp &= \lbr n n c\up{\dagger}_{\ell'\beta} c_{\ell'\beta} \rbr = M' \mathcal{I}_{3,0} + \mathcal{I}_{3,1} \,, \\ 
\Tcpp &= \lbr n c\up{\dagger}_{\ell'\beta} n c_{\ell'\beta} \rbr = M'(\mathcal{I}_{3,0}-\mathcal{I}_{1,0}) + \mathcal{I}_{3,1}-\mathcal{I}_{1,1} \,, \\
\Tappp &= \lbr n \nt \nt n \rbr =  \mathcal{I}_{4,0} - 2\nf\mathcal{I}_{3,0} + \nf\up{2}\mathcal{I}_{2,0} \,, \\
\Tbppp &= \lbr n n \nt \nt \rbr =  \mathcal{I}_{4,0} - 2\nf\mathcal{I}_{3,0} + \nf\up{2}\mathcal{I}_{2,0} \,, \\
\Tcppp &= \lbr n \nt n \nt \rbr =  \mathcal{I}_{4,0} - 2\nf\mathcal{I}_{3,0} + \nf\up{2}\mathcal{I}_{2,0} \,.
\end{align}
\endgroup
Using Eqs. \ref{eq:d} and \ref{eq:n4} we have,
\begin{align}
\langle O_{4} \rangle = \frac{N_{4}}{D} &= \To \bigg \lbrace 
1 + \gam\up{2} \left[ \left( \frac{\Ta}{\To} - \lo \right) \Da 
+ \left( \frac{\Tb}{\To} - \lo \right) \Db 
+\left( \frac{\Tc}{\To} - \lo \right) \Dc \right]  \nonumber \\
&+ \gc\up{2} \left[ \left( \frac{\Tap}{\To} - \lop \right) \Dap 
+ \left( \frac{\Tbp}{\To} - \lop \right) \Dbp 
+\left( \frac{\Tcp}{\To} - \lop \right) \Dcp \right] \nonumber \\
&+ \gc\up{2} \left[ \left( \frac{\Tapp}{\To} - \lopp \right) \Dapp 
+ \left( \frac{\Tbpp}{\To} - \lopp \right) \Dbpp 
+\left( \frac{\Tcpp}{\To} - \lopp \right) \Dcpp \right] \nonumber \\
&+ \gv\up{2} \left[ \left( \frac{\Tappp}{\To} - \loppp \right) \Daz 
+ \left( \frac{\Tbppp}{\To} - \loppp \right) \Dbz 
+\left( \frac{\Tcppp}{\To} - \loppp \right) \Dcz \right]
\bigg \rbrace \,.
\end{align}
Therefore, we obtain, 
\begin{equation}
Z_{n} = Z_{\nt} = 1 - \frac{\gamr\up{2}}{\ep} \Tgam - \frac{\gcr\up{2}}{2\rb} \Tg - \frac{\gvr\up{2}}{\ep} \Tv \,,
\end{equation}
where
\begin{align}
\Tgam &= \frac{\Ta + \Tb -2\Tc}{\To} \,, \\
\Tg &= \frac{\Tap + \Tbp - 2\Tcp + \Tapp + \Tbpp - 2\Tcpp}{\To} \,, \\
\Tv &= \frac{\Tappp + \Tbppp -2\Tcppp}{\To} \,.
\end{align}
We find that $\Tg=2$, $\Tgam=0$ and $\Tv=0$ for $M=2\,, M'=1$, . Thus, for $M=2\,, M'=1$,
\begin{equation}
\label{eq:zn_m}
Z_{n} = Z_{\nt} = 1 - \frac{\gcr\up{2}}{\rb} \,.
\end{equation}

\subsection{Beta functions}
\label{sec:beta_fp}

With the renormalization factors for the gauge-invariant operators at hand, we can obtain the beta functions in a straightforward manner. Note that due to the absence of interaction terms the renormalization factors for the coupling constants are all unity, i.e., $\widetilde{Z}_{\gcr}=\widetilde{Z}_{\gamr}=\widetilde{Z}_{\gvr}=1$. Now using Eq. \ref{eq:renorm_fact_m} we find,
\begin{align}
\label{eq:beta1_m}
\frac{\ep}{2}\gamr Z_{S} + \left[ Z_{S} - \frac{\gamr}{2} \frac{\partial Z_{S}}{\partial \gamr} \right] \betgam 
- \frac{\gamr}{2} \frac{\partial Z_{S}}{\partial \gcr} \betg - \frac{\gamr}{2} \frac{\partial Z_{S}}{\partial \gvr} \betv &= 0 
\,, \\
\label{eq:beta2_m}
\rb\gcr Z_{c} + \left[ Z_{c} - \frac{\gcr}{2} \frac{\partial Z_{c}}{\partial \gcr} \right] \betg 
- \frac{\gcr}{2} \frac{\partial Z_{c}}{\partial \gamr} \betgam 
- \frac{\gcr}{2} \frac{\partial Z_{c}}{\partial \gv} \betv &= 0 \,, \\
\label{eq:beta3_m}
\frac{\epp}{2}\gvr Z_{\nt} + \left[ Z_{\nt} - \frac{\gvr}{2} \frac{\partial Z_{v}}{\partial \gvr} \right] \betv 
- \frac{\gvr}{2} \frac{\partial Z_{\nt}}{\partial \gcr} \betg 
- \frac{\gvr}{2} \frac{\partial Z_{\nt}}{\partial \gamr} \betgam &= 0 \,.
\end{align}
We now solve the above three equations using Eqs. \ref{eq:zs_m}, \ref{eq:zc_m} and \ref{eq:zn_m}, and obtain the one-loop beta functions,
\begin{align}
\label{eq:beta_g_m}
\betg &= -\rb \gcr + \frac{3}{2} \gcr\up{3} + \frac{3}{8} \gcr \gamr\up{2} + \frac{1}{2} \gcr \gvr\up{2} \,, \\
\label{eq:beta_gam_m}
\betgam &= -\frac{\ep}{2} \gamr + \gamr\up{3} + \gcr\up{2} \gamr \,, \\
\label{eq:beta_v_m}
\betv &= -\frac{\epp}{2} \gvr + \gcr\up{2} \gvr \,.
\end{align}
These are exactly the same as obtained earlier via a different RG procedure in Sec. \ref{sec:beta}. The calculation of the rest of the details such as the fixed points and anomalous dimensions follow exactly as discussed in the main text.

\section{Supersymmetry}
\label{app:bath}

In this appendix, we explore the possibility that averaged Hamiltonians 
$H_{\rm imp} + H_{\rm bath}$ in (\ref{eq:H_imp}) exhibit $SU(1|2)$ supersymmetry. We were unable to define a suitable 
supersymmetry operation, as we discuss below. The difficult lies in making the bath supersymmetric. One approach is try to implement a spacetime supersymmetry on the bath fermions $\psi_\alpha$ and the bosons $\phi$ and $\zeta$: however that does not work because the scaling dimensions of fermions and bosons are not equal in this supersymmetry, whereas equality of the power-laws in (\ref{QRpower}) requires them to have the same scaling dimensions. 

More progress is possible in an approach which fractionalizes the bath operators, in a manner which parallels the impurity site. So we write
%
\bea
\psi_\alpha (0) &=& \frac{1}{\Omega} \sum_k \widetilde{f}_{k\alpha} \widetilde{b}_k^\dagger \nonumber\\
\phi_a (0) &=& \frac{1}{\Omega} \sum_k \widetilde{f}_{k\alpha}^\dagger \frac{\sigma^a_{\alpha\beta}}{2} \widetilde{f}_{k\beta}  \nonumber \\
\zeta (0) &=& \frac{1}{\Omega} \sum_k \widetilde{f}_{k\alpha}^\dagger \widetilde{f}_{k\alpha} \,,
\eea
where $\Omega$ is a suitable normalization of the sum over $k$.
The Green's functions of the partons
\bea
\widetilde{G}_f (k,\tau) \, \delta_{\alpha\beta} &=& - \left\langle  \widetilde{f}_{k\alpha}(\tau) \widetilde{f}_{k\beta}^\dagger (0)  \right\rangle \nonumber \\
\widetilde{G}_b (k,\tau) &=& - \left\langle  \widetilde{b}_k (\tau) \widetilde{b}_k^\dagger (0)  \right\rangle \,,
\eea
can then be used to obtain the fields in (\ref{eq:self_cons})
\bea
R(\tau) &=& - \frac{1}{\Omega} \sum_k \widetilde{G}_f (k,\tau) \widetilde{G}_b (k,-\tau) \nonumber \\
Q(\tau) &=& -  \frac{1}{2\Omega} \sum_k \widetilde{G}_f (k,\tau) \widetilde{G}_f (k,-\tau) \nonumber \\
P(\tau) &=& - \frac{2}{\Omega} \sum_k \widetilde{G}_f (k,\tau) \widetilde{G}_f (k,-\tau) \,.
\eea
Finally, we replace the bath Hamiltonian in (\ref{eq:H_imp}) by
\beq
\widetilde{H}_{\rm bath} = \frac{1}{\Omega} \sum_k \epsilon_f (k) \widetilde{f}_{k \alpha}^\dagger \widetilde{f}_{k \alpha} + \frac{1}{\Omega} \sum_k \epsilon_b (k) \widetilde{b}_{k}^\dagger \widetilde{b}_{k}\,.
\eeq
Now we consider generators of the $SU(1|2)$ superalgebra as the sum of impurity and bath terms, replacing (\ref{fractionalize},\ref{super}) by
\bea
\mathcal{C}_\alpha &=& f_{\alpha} b^\dagger + \frac{1}{\Omega} \sum_k \widetilde{f}_{k \alpha} \widetilde{b}_k^\dagger \nonumber \\
\mathcal{S}^a &=& f_\alpha^\dagger \frac{\sigma^a_{\alpha\beta}}{2} f_\beta + \frac{1}{\Omega} \sum_k \widetilde{f}_{k \alpha}^\dagger\frac{\sigma^a_{\alpha\beta}}{2} \widetilde{f}_{k \beta} \nonumber \\
\mathcal{V} &=& \frac{1}{2} f_\alpha^\dagger f_\alpha + b^\dagger b + \frac{1}{2\Omega} \sum_k \widetilde{f}_{k \alpha}^\dagger \widetilde{f}_{k \alpha} + \frac{1}{\Omega} \sum_k \widetilde{b}_k^\dagger \widetilde{b}_k
\eea
It is now easy to see that $H_{\rm imp}$ and $\widetilde{H}_{\rm bath}$ both commute with $\mathcal{S}^a$ and $\mathcal{V}$. We can also find by explicit evaluation that
\bea
\left[ \mathcal{C}_\alpha, H_{\rm bath} \right] &=& 0\,, \quad \mbox{for $\epsilon_f (k) = \epsilon_b (k)$}\,. \label{CH}
\eea
Further, 
\begin{align}
\left[ \mathcal{C}_{\alpha}, H_{\rm imp} \right] &=  (s_{0} + \lam) c_{\alpha} - \lam c_{\alpha} 
+ \gc (\delta_{\alpha \beta} V + \sigma\up{a}_{\alpha \beta} S\up{a}) \psi_{\beta}(0) 
+\gc (\delta_{\alpha \beta} \widetilde{V} + \sigma\up{a}_{\alpha \beta} \phi_{a} (0)) c_{\beta} \nonumber \\ 
&+\gam (\frac{\sigma\up{a}_{\alpha \beta}}{2}  c_{\beta} \phi_{a}(0) + \frac{\sigma\up{a}_{\alpha \beta}}{2} S\up{a} \psi_{\beta}(0)) 
+ \gv (c_{\alpha} \zeta(0) + f\up{\dagger}_{\beta} f_{\beta} \psi_{\alpha}(0)) - n_{f} \gv \psi_{\alpha} (0) \,,
\end{align}
where $\widetilde{V} = (1/\Omega) \sum_{k} (\widetilde{f}\up{\dagger}_{k\alpha} \widetilde{f}_{k\alpha}/2 + \widetilde{b}\up{\dagger}_{k} \widetilde{b}_{k})$. Now, recall that $f\up{\dagger}_{\beta} f_{\beta} = 2 - 2V$, using Eq. \ref{fractionalize} and the constraint $f\up{\dagger}_{\beta} f_{\beta} + b\up{\dagger} b =1$. For the bath operators we include a chemical potential such that  $(1/\Omega) \sum_{k} (\widetilde{f}\up{\dagger}_{k\beta} \widetilde{f}_{k\beta} + \widetilde{b}\up{\dagger}_k \widetilde{b}_k) =1$; then one can write $\zeta(0) = 2 - 2\widetilde{V}$. In this case, for $s_{0}=-n_{f} \gv$, $\gam =-2\gc$, and $\gc=2\gv$ we obtain,
\bea
 \left[ \mathcal{C}_\alpha, H_{\rm imp} \right] &=& s_0 \, \mathcal{C}_\alpha \,, 
\eea
which is similar to (\ref{cHtJK}).

However, the condition in (\ref{CH}) leads to an issue with supersymmetry in the class of models studied in the body of the paper. To obtain the ansatz in (\ref{QRpower}), with $R(\tau)$ an odd function of $\tau$ and $P(\tau),Q(\tau)$ even functions of $\tau$, we need $\epsilon_f (k)$ to be an odd function of $k$, while $\epsilon_b (k)$ needs to be positive for stability. This is incompatible with the requirements of supersymmetry.

\bibliography{randomtJV_rg}

\begin{thebibliography}{26}%
\makeatletter
\providecommand \@ifxundefined [1]{%
 \@ifx{#1\undefined}
}%
\providecommand \@ifnum [1]{%
 \ifnum #1\expandafter \@firstoftwo
 \else \expandafter \@secondoftwo
 \fi
}%
\providecommand \@ifx [1]{%
 \ifx #1\expandafter \@firstoftwo
 \else \expandafter \@secondoftwo
 \fi
}%
\providecommand \natexlab [1]{#1}%
\providecommand \enquote  [1]{``#1''}%
\providecommand \bibnamefont  [1]{#1}%
\providecommand \bibfnamefont [1]{#1}%
\providecommand \citenamefont [1]{#1}%
\providecommand \href@noop [0]{\@secondoftwo}%
\providecommand \href [0]{\begingroup \@sanitize@url \@href}%
\providecommand \@href[1]{\@@startlink{#1}\@@href}%
\providecommand \@@href[1]{\endgroup#1\@@endlink}%
\providecommand \@sanitize@url [0]{\catcode `\\12\catcode `\$12\catcode
  `\&12\catcode `\#12\catcode `\^12\catcode `\_12\catcode `\%12\relax}%
\providecommand \@@startlink[1]{}%
\providecommand \@@endlink[0]{}%
\providecommand \url  [0]{\begingroup\@sanitize@url \@url }%
\providecommand \@url [1]{\endgroup\@href {#1}{\urlprefix }}%
\providecommand \urlprefix  [0]{URL }%
\providecommand \Eprint [0]{\href }%
\providecommand \doibase [0]{http://dx.doi.org/}%
\providecommand \selectlanguage [0]{\@gobble}%
\providecommand \bibinfo  [0]{\@secondoftwo}%
\providecommand \bibfield  [0]{\@secondoftwo}%
\providecommand \translation [1]{[#1]}%
\providecommand \BibitemOpen [0]{}%
\providecommand \bibitemStop [0]{}%
\providecommand \bibitemNoStop [0]{.\EOS\space}%
\providecommand \EOS [0]{\spacefactor3000\relax}%
\providecommand \BibitemShut  [1]{\csname bibitem#1\endcsname}%
\let\auto@bib@innerbib\@empty
\bibitem [{\citenamefont {{He}}\ \emph {et~al.}(2019)\citenamefont {{He}},
  \citenamefont {{Rotundu}}, \citenamefont {{Scheurer}}, \citenamefont {{He}},
  \citenamefont {{Hashimoto}}, \citenamefont {{Xu}}, \citenamefont {{Wang}},
  \citenamefont {{Huang}}, \citenamefont {{Jia}}, \citenamefont {{Chen}},
  \citenamefont {{Moritz}}, \citenamefont {{Lu}}, \citenamefont {{Lee}},
  \citenamefont {{Devereaux}},\ and\ \citenamefont {{Shen}}}]{Shen18}%
  \BibitemOpen
  \bibfield  {author} {\bibinfo {author} {\bibfnamefont {J.-F.}\ \bibnamefont
  {{He}}}, \bibinfo {author} {\bibfnamefont {C.~R.}\ \bibnamefont {{Rotundu}}},
  \bibinfo {author} {\bibfnamefont {M.~S.}\ \bibnamefont {{Scheurer}}},
  \bibinfo {author} {\bibfnamefont {Y.}~\bibnamefont {{He}}}, \bibinfo {author}
  {\bibfnamefont {M.}~\bibnamefont {{Hashimoto}}}, \bibinfo {author}
  {\bibfnamefont {K.}~\bibnamefont {{Xu}}}, \bibinfo {author} {\bibfnamefont
  {Y.}~\bibnamefont {{Wang}}}, \bibinfo {author} {\bibfnamefont {E.~W.}\
  \bibnamefont {{Huang}}}, \bibinfo {author} {\bibfnamefont {T.}~\bibnamefont
  {{Jia}}}, \bibinfo {author} {\bibfnamefont {S.-D.}\ \bibnamefont {{Chen}}},
  \bibinfo {author} {\bibfnamefont {B.}~\bibnamefont {{Moritz}}}, \bibinfo
  {author} {\bibfnamefont {D.-H.}\ \bibnamefont {{Lu}}}, \bibinfo {author}
  {\bibfnamefont {Y.~S.}\ \bibnamefont {{Lee}}}, \bibinfo {author}
  {\bibfnamefont {T.~P.}\ \bibnamefont {{Devereaux}}}, \ and\ \bibinfo {author}
  {\bibfnamefont {Z.-X.}\ \bibnamefont {{Shen}}},\ }\bibfield  {title}
  {\enquote {\bibinfo {title} {{Fermi surface reconstruction in electron-doped
  cuprates without antiferromagnetic long-range order}},}\ }\href {\doibase
  10.1073/pnas.1816121116} {\bibfield  {journal} {\bibinfo  {journal} {Proc.
  Nat. Acad. Sci.}\ ,\ \bibinfo {pages} {online}} (\bibinfo {year} {2019})},\
  \Eprint {http://arxiv.org/abs/1811.04992} {arXiv:1811.04992
  [cond-mat.supr-con]} \BibitemShut {NoStop}%
\bibitem [{\citenamefont {Chen}\ \emph {et~al.}(2019)\citenamefont {Chen},
  \citenamefont {Hashimoto}, \citenamefont {He}, \citenamefont {Song},
  \citenamefont {Xu}, \citenamefont {He}, \citenamefont {Devereaux},
  \citenamefont {Eisaki}, \citenamefont {Lu}, \citenamefont {Zaanen},\ and\
  \citenamefont {Shen}}]{Shen19}%
  \BibitemOpen
  \bibfield  {author} {\bibinfo {author} {\bibfnamefont {S.-D.}\ \bibnamefont
  {Chen}}, \bibinfo {author} {\bibfnamefont {M.}~\bibnamefont {Hashimoto}},
  \bibinfo {author} {\bibfnamefont {Y.}~\bibnamefont {He}}, \bibinfo {author}
  {\bibfnamefont {D.}~\bibnamefont {Song}}, \bibinfo {author} {\bibfnamefont
  {K.-J.}\ \bibnamefont {Xu}}, \bibinfo {author} {\bibfnamefont {J.-F.}\
  \bibnamefont {He}}, \bibinfo {author} {\bibfnamefont {T.~P.}\ \bibnamefont
  {Devereaux}}, \bibinfo {author} {\bibfnamefont {H.}~\bibnamefont {Eisaki}},
  \bibinfo {author} {\bibfnamefont {D.-H.}\ \bibnamefont {Lu}}, \bibinfo
  {author} {\bibfnamefont {J.}~\bibnamefont {Zaanen}}, \ and\ \bibinfo {author}
  {\bibfnamefont {Z.-X.}\ \bibnamefont {Shen}},\ }\bibfield  {title} {\enquote
  {\bibinfo {title} {{Incoherent strange metal sharply bounded by a critical
  doping in Bi2212}},}\ }\href {\doibase 10.1126/science.aaw8850} {\bibfield
  {journal} {\bibinfo  {journal} {Science}\ }\textbf {\bibinfo {volume}
  {366}},\ \bibinfo {pages} {1099} (\bibinfo {year} {2019})}\BibitemShut
  {NoStop}%
\bibitem [{\citenamefont {{Michon}}\ \emph {et~al.}(2019)\citenamefont
  {{Michon}}, \citenamefont {{Girod}}, \citenamefont {{Badoux}}, \citenamefont
  {{Ka{\v{c}}mar{\v{c}}{\'\i}k}}, \citenamefont {{Ma}}, \citenamefont
  {{Dragomir}}, \citenamefont {{Dabkowska}}, \citenamefont {{Gaulin}},
  \citenamefont {{Zhou}}, \citenamefont {{Pyon}}, \citenamefont {{Takayama}},
  \citenamefont {{Takagi}}, \citenamefont {{Verret}}, \citenamefont
  {{Doiron-Leyraud}}, \citenamefont {{Marcenat}}, \citenamefont {{Taillefer}},\
  and\ \citenamefont {{Klein}}}]{Michon18}%
  \BibitemOpen
  \bibfield  {author} {\bibinfo {author} {\bibfnamefont {B.}~\bibnamefont
  {{Michon}}}, \bibinfo {author} {\bibfnamefont {C.}~\bibnamefont {{Girod}}},
  \bibinfo {author} {\bibfnamefont {S.}~\bibnamefont {{Badoux}}}, \bibinfo
  {author} {\bibfnamefont {J.}~\bibnamefont {{Ka{\v{c}}mar{\v{c}}{\'\i}k}}},
  \bibinfo {author} {\bibfnamefont {Q.}~\bibnamefont {{Ma}}}, \bibinfo {author}
  {\bibfnamefont {M.}~\bibnamefont {{Dragomir}}}, \bibinfo {author}
  {\bibfnamefont {H.~A.}\ \bibnamefont {{Dabkowska}}}, \bibinfo {author}
  {\bibfnamefont {B.~D.}\ \bibnamefont {{Gaulin}}}, \bibinfo {author}
  {\bibfnamefont {J.~S.}\ \bibnamefont {{Zhou}}}, \bibinfo {author}
  {\bibfnamefont {S.}~\bibnamefont {{Pyon}}}, \bibinfo {author} {\bibfnamefont
  {T.}~\bibnamefont {{Takayama}}}, \bibinfo {author} {\bibfnamefont
  {H.}~\bibnamefont {{Takagi}}}, \bibinfo {author} {\bibfnamefont
  {S.}~\bibnamefont {{Verret}}}, \bibinfo {author} {\bibfnamefont
  {N.}~\bibnamefont {{Doiron-Leyraud}}}, \bibinfo {author} {\bibfnamefont
  {C.}~\bibnamefont {{Marcenat}}}, \bibinfo {author} {\bibfnamefont
  {L.}~\bibnamefont {{Taillefer}}}, \ and\ \bibinfo {author} {\bibfnamefont
  {T.}~\bibnamefont {{Klein}}},\ }\bibfield  {title} {\enquote {\bibinfo
  {title} {{Thermodynamic signatures of quantum criticality in cuprate
  superconductors}},}\ }\href {\doibase 10.1038/s41586-019-0932-x} {\bibfield
  {journal} {\bibinfo  {journal} {Nature}\ }\textbf {\bibinfo {volume} {567}},\
  \bibinfo {pages} {218} (\bibinfo {year} {2019})},\ \Eprint
  {http://arxiv.org/abs/1804.08502} {arXiv:1804.08502 [cond-mat.supr-con]}
  \BibitemShut {NoStop}%
\bibitem [{\citenamefont {{Mitrano}}\ \emph {et~al.}(2018)\citenamefont
  {{Mitrano}}, \citenamefont {{Husain}}, \citenamefont {{Vig}}, \citenamefont
  {{Kogar}}, \citenamefont {{Rak}}, \citenamefont {{Rubeck}}, \citenamefont
  {{Schmalian}}, \citenamefont {{Uchoa}}, \citenamefont {{Schneeloch}},
  \citenamefont {{Zhong}}, \citenamefont {{Gu}},\ and\ \citenamefont
  {{Abbamonte}}}]{Mitrano18}%
  \BibitemOpen
  \bibfield  {author} {\bibinfo {author} {\bibfnamefont {M.}~\bibnamefont
  {{Mitrano}}}, \bibinfo {author} {\bibfnamefont {A.~A.}\ \bibnamefont
  {{Husain}}}, \bibinfo {author} {\bibfnamefont {S.}~\bibnamefont {{Vig}}},
  \bibinfo {author} {\bibfnamefont {A.}~\bibnamefont {{Kogar}}}, \bibinfo
  {author} {\bibfnamefont {M.~S.}\ \bibnamefont {{Rak}}}, \bibinfo {author}
  {\bibfnamefont {S.~I.}\ \bibnamefont {{Rubeck}}}, \bibinfo {author}
  {\bibfnamefont {J.}~\bibnamefont {{Schmalian}}}, \bibinfo {author}
  {\bibfnamefont {B.}~\bibnamefont {{Uchoa}}}, \bibinfo {author} {\bibfnamefont
  {J.}~\bibnamefont {{Schneeloch}}}, \bibinfo {author} {\bibfnamefont
  {R.}~\bibnamefont {{Zhong}}}, \bibinfo {author} {\bibfnamefont {G.~D.}\
  \bibnamefont {{Gu}}}, \ and\ \bibinfo {author} {\bibfnamefont
  {P.}~\bibnamefont {{Abbamonte}}},\ }\bibfield  {title} {\enquote {\bibinfo
  {title} {{Anomalous density fluctuations in a strange metal}},}\ }\href
  {\doibase 10.1073/pnas.1721495115} {\bibfield  {journal} {\bibinfo  {journal}
  {Proceedings of the National Academy of Science}\ }\textbf {\bibinfo {volume}
  {115}},\ \bibinfo {pages} {5392} (\bibinfo {year} {2018})},\ \Eprint
  {http://arxiv.org/abs/1708.01929} {arXiv:1708.01929 [cond-mat.str-el]}
  \BibitemShut {NoStop}%
\bibitem [{\citenamefont {{Husain}}\ \emph {et~al.}(2019)\citenamefont
  {{Husain}}, \citenamefont {{Mitrano}}, \citenamefont {{Rak}}, \citenamefont
  {{Rubeck}}, \citenamefont {{Uchoa}}, \citenamefont {{Schneeloch}},
  \citenamefont {{Zhong}}, \citenamefont {{Gu}},\ and\ \citenamefont
  {{Abbamonte}}}]{Husain19}%
  \BibitemOpen
  \bibfield  {author} {\bibinfo {author} {\bibfnamefont {A.~A.}\ \bibnamefont
  {{Husain}}}, \bibinfo {author} {\bibfnamefont {M.}~\bibnamefont {{Mitrano}}},
  \bibinfo {author} {\bibfnamefont {M.~S.}\ \bibnamefont {{Rak}}}, \bibinfo
  {author} {\bibfnamefont {S.~I.}\ \bibnamefont {{Rubeck}}}, \bibinfo {author}
  {\bibfnamefont {B.}~\bibnamefont {{Uchoa}}}, \bibinfo {author} {\bibfnamefont
  {J.}~\bibnamefont {{Schneeloch}}}, \bibinfo {author} {\bibfnamefont
  {R.}~\bibnamefont {{Zhong}}}, \bibinfo {author} {\bibfnamefont {G.~D.}\
  \bibnamefont {{Gu}}}, \ and\ \bibinfo {author} {\bibfnamefont
  {P.}~\bibnamefont {{Abbamonte}}},\ }\bibfield  {title} {\enquote {\bibinfo
  {title} {{Crossover of Charge Fluctuations across the Strange Metal Phase
  Diagram}},}\ }\href {\doibase 10.1103/PhysRevX.9.041062} {\bibfield
  {journal} {\bibinfo  {journal} {Phys. Rev. X}\ }\textbf {\bibinfo {volume}
  {9}},\ \bibinfo {pages} {041062} (\bibinfo {year} {2019})},\ \Eprint
  {http://arxiv.org/abs/1903.04038} {arXiv:1903.04038 [cond-mat.str-el]}
  \BibitemShut {NoStop}%
\bibitem [{\citenamefont {{Husain}}\ \emph {et~al.}()\citenamefont {{Husain}},
  \citenamefont {{Mitrano}}, \citenamefont {{Rak}}, \citenamefont {{Rubeck}},
  \citenamefont {{Yang}}, \citenamefont {{Sow}}, \citenamefont {{Maeno}},
  \citenamefont {{Batson}},\ and\ \citenamefont {{Abbamonte}}}]{Husain20}%
  \BibitemOpen
  \bibfield  {author} {\bibinfo {author} {\bibfnamefont {A.~A.}\ \bibnamefont
  {{Husain}}}, \bibinfo {author} {\bibfnamefont {M.}~\bibnamefont {{Mitrano}}},
  \bibinfo {author} {\bibfnamefont {M.~S.}\ \bibnamefont {{Rak}}}, \bibinfo
  {author} {\bibfnamefont {S.~I.}\ \bibnamefont {{Rubeck}}}, \bibinfo {author}
  {\bibfnamefont {H.}~\bibnamefont {{Yang}}}, \bibinfo {author} {\bibfnamefont
  {C.}~\bibnamefont {{Sow}}}, \bibinfo {author} {\bibfnamefont
  {Y.}~\bibnamefont {{Maeno}}}, \bibinfo {author} {\bibfnamefont {P.~E.}\
  \bibnamefont {{Batson}}}, \ and\ \bibinfo {author} {\bibfnamefont
  {P.}~\bibnamefont {{Abbamonte}}},\ }\bibfield  {title} {\enquote {\bibinfo
  {title} {{Coexisting Fermi Liquid and Strange Metal Phenomena in
  Sr$_2$RuO$_4$}},}\ }\href@noop {} {\ }\Eprint
  {http://arxiv.org/abs/2007.06670} {arXiv:2007.06670 [cond-mat.str-el]}
  \BibitemShut {NoStop}%
\bibitem [{\citenamefont {Joshi}\ \emph {et~al.}(2020)\citenamefont {Joshi},
  \citenamefont {Li}, \citenamefont {Tarnopolsky}, \citenamefont {Georges},\
  and\ \citenamefont {Sachdev}}]{dqcp_tj}%
  \BibitemOpen
  \bibfield  {author} {\bibinfo {author} {\bibfnamefont {D.~G.}\ \bibnamefont
  {Joshi}}, \bibinfo {author} {\bibfnamefont {C.}~\bibnamefont {Li}}, \bibinfo
  {author} {\bibfnamefont {G.}~\bibnamefont {Tarnopolsky}}, \bibinfo {author}
  {\bibfnamefont {A.}~\bibnamefont {Georges}}, \ and\ \bibinfo {author}
  {\bibfnamefont {S.}~\bibnamefont {Sachdev}},\ }\bibfield  {title} {\enquote
  {\bibinfo {title} {{Deconfined critical point in a doped random quantum
  Heisenberg magnet}},}\ }\href {\doibase 10.1103/PhysRevX.10.021033}
  {\bibfield  {journal} {\bibinfo  {journal} {Phys. Rev. X}\ }\textbf {\bibinfo
  {volume} {10}},\ \bibinfo {pages} {021033} (\bibinfo {year} {2020})},\
  \Eprint {http://arxiv.org/abs/1912.08822} {arXiv:1912.08822
  [cond-mat.str-el]} \BibitemShut {NoStop}%
\bibitem [{\citenamefont {{Sachdev}}\ and\ \citenamefont {{Ye}}(1993)}]{SY92}%
  \BibitemOpen
  \bibfield  {author} {\bibinfo {author} {\bibfnamefont {S.}~\bibnamefont
  {{Sachdev}}}\ and\ \bibinfo {author} {\bibfnamefont {J.}~\bibnamefont
  {{Ye}}},\ }\bibfield  {title} {\enquote {\bibinfo {title} {{Gapless
  spin-fluid ground state in a random quantum Heisenberg magnet}},}\ }\href
  {\doibase 10.1103/PhysRevLett.70.3339} {\bibfield  {journal} {\bibinfo
  {journal} {Phys. Rev. Lett.}\ }\textbf {\bibinfo {volume} {70}},\ \bibinfo
  {pages} {3339} (\bibinfo {year} {1993})},\ \Eprint
  {http://arxiv.org/abs/cond-mat/9212030} {cond-mat/9212030} \BibitemShut
  {NoStop}%
\bibitem [{\citenamefont {{Kitaev}}(2015)}]{kitaev2015talk}%
  \BibitemOpen
  \bibfield  {author} {\bibinfo {author} {\bibfnamefont {A.~Y.}\ \bibnamefont
  {{Kitaev}}},\ }\bibfield  {title} {\enquote {\bibinfo {title} {{Talks at
  KITP, University of California, Santa Barbara}},}\ }\href
  {http://online.kitp.ucsb.edu/online/entangled15/} {\bibfield  {journal}
  {\bibinfo  {journal} {Entanglement in Strongly-Correlated Quantum Matter}\ }
  (\bibinfo {year} {2015})}\BibitemShut {NoStop}%
\bibitem [{\citenamefont {Wiegmann}(1988)}]{Wiegmann88}%
  \BibitemOpen
  \bibfield  {author} {\bibinfo {author} {\bibfnamefont {P.~B.}\ \bibnamefont
  {Wiegmann}},\ }\bibfield  {title} {\enquote {\bibinfo {title}
  {Superconductivity in strongly correlated electronic systems and confinement
  versus deconfinement phenomenon},}\ }\href {\doibase
  10.1103/PhysRevLett.60.821} {\bibfield  {journal} {\bibinfo  {journal} {Phys.
  Rev. Lett.}\ }\textbf {\bibinfo {volume} {60}},\ \bibinfo {pages} {821}
  (\bibinfo {year} {1988})}\BibitemShut {NoStop}%
\bibitem [{\citenamefont {F\"orster}(1989)}]{Forster89}%
  \BibitemOpen
  \bibfield  {author} {\bibinfo {author} {\bibfnamefont {D.}~\bibnamefont
  {F\"orster}},\ }\bibfield  {title} {\enquote {\bibinfo {title} {Staggered
  spin and statistics in the supersymmetric t-j model},}\ }\href {\doibase
  10.1103/PhysRevLett.63.2140} {\bibfield  {journal} {\bibinfo  {journal}
  {Phys. Rev. Lett.}\ }\textbf {\bibinfo {volume} {63}},\ \bibinfo {pages}
  {2140} (\bibinfo {year} {1989})}\BibitemShut {NoStop}%
\bibitem [{\citenamefont {Bares}\ and\ \citenamefont
  {Blatter}(1990)}]{Blatter90}%
  \BibitemOpen
  \bibfield  {author} {\bibinfo {author} {\bibfnamefont {P.~A.}\ \bibnamefont
  {Bares}}\ and\ \bibinfo {author} {\bibfnamefont {G.}~\bibnamefont
  {Blatter}},\ }\bibfield  {title} {\enquote {\bibinfo {title} {{Supersymmetric
  $t$-$J$ model in one dimension: Separation of spin and charge}},}\ }\href
  {\doibase 10.1103/PhysRevLett.64.2567} {\bibfield  {journal} {\bibinfo
  {journal} {Phys. Rev. Lett.}\ }\textbf {\bibinfo {volume} {64}},\ \bibinfo
  {pages} {2567} (\bibinfo {year} {1990})}\BibitemShut {NoStop}%
\bibitem [{\citenamefont {{G{\"o}hmann}}\ and\ \citenamefont
  {{Seel}}(2003)}]{Czech03}%
  \BibitemOpen
  \bibfield  {author} {\bibinfo {author} {\bibfnamefont {F.}~\bibnamefont
  {{G{\"o}hmann}}}\ and\ \bibinfo {author} {\bibfnamefont {A.}~\bibnamefont
  {{Seel}}},\ }\bibfield  {title} {\enquote {\bibinfo {title} {{A Note on the
  Bethe Ansatz Solution of the Supersymmetric $t$-$J$ Model}},}\ }\href
  {\doibase 10.1023/B:CJOP.0000010530.54520.12} {\bibfield  {journal} {\bibinfo
   {journal} {Czechoslovak Journal of Physics}\ }\textbf {\bibinfo {volume}
  {53}},\ \bibinfo {pages} {1041} (\bibinfo {year} {2003})},\ \Eprint
  {http://arxiv.org/abs/cond-mat/0309138} {arXiv:cond-mat/0309138
  [cond-mat.str-el]} \BibitemShut {NoStop}%
\bibitem [{\citenamefont {Sarkar}(1991)}]{Sarkar91}%
  \BibitemOpen
  \bibfield  {author} {\bibinfo {author} {\bibfnamefont {S.}~\bibnamefont
  {Sarkar}},\ }\bibfield  {title} {\enquote {\bibinfo {title} {{The
  supersymmetric $t$-$J$ model in one dimension}},}\ }\href {\doibase
  10.1088/0305-4470/24/5/026} {\bibfield  {journal} {\bibinfo  {journal} {J.
  Phys. A: Math. Gen.}\ }\textbf {\bibinfo {volume} {24}},\ \bibinfo {pages}
  {1137} (\bibinfo {year} {1991})}\BibitemShut {NoStop}%
\bibitem [{\citenamefont {Essler}\ and\ \citenamefont
  {Korepin}(1992)}]{Essler92}%
  \BibitemOpen
  \bibfield  {author} {\bibinfo {author} {\bibfnamefont {F.~H.~L.}\
  \bibnamefont {Essler}}\ and\ \bibinfo {author} {\bibfnamefont {V.~E.}\
  \bibnamefont {Korepin}},\ }\bibfield  {title} {\enquote {\bibinfo {title}
  {{Higher conservation laws and algebraic Bethe Ans\"atze for the
  supersymmetric $t$-$J$ model}},}\ }\href {\doibase 10.1103/PhysRevB.46.9147}
  {\bibfield  {journal} {\bibinfo  {journal} {Phys. Rev. B}\ }\textbf {\bibinfo
  {volume} {46}},\ \bibinfo {pages} {9147} (\bibinfo {year}
  {1992})}\BibitemShut {NoStop}%
\bibitem [{\citenamefont {{Sachdev}}\ \emph {et~al.}(1999)\citenamefont
  {{Sachdev}}, \citenamefont {{Buragohain}},\ and\ \citenamefont
  {{Vojta}}}]{SBV1999}%
  \BibitemOpen
  \bibfield  {author} {\bibinfo {author} {\bibfnamefont {S.}~\bibnamefont
  {{Sachdev}}}, \bibinfo {author} {\bibfnamefont {C.}~\bibnamefont
  {{Buragohain}}}, \ and\ \bibinfo {author} {\bibfnamefont {M.}~\bibnamefont
  {{Vojta}}},\ }\bibfield  {title} {\enquote {\bibinfo {title} {{Quantum
  Impurity in a Nearly Critical Two Dimensional Antiferromagnet}},}\ }\href
  {\doibase 10.1126/science.286.5449.2479} {\bibfield  {journal} {\bibinfo
  {journal} {Science}\ }\textbf {\bibinfo {volume} {286}},\ \bibinfo {pages}
  {2479} (\bibinfo {year} {1999})},\ \Eprint
  {http://arxiv.org/abs/cond-mat/0004156} {arXiv:cond-mat/0004156
  [cond-mat.str-el]} \BibitemShut {NoStop}%
\bibitem [{\citenamefont {{Vojta}}\ \emph {et~al.}(2000)\citenamefont
  {{Vojta}}, \citenamefont {{Buragohain}},\ and\ \citenamefont
  {{Sachdev}}}]{VBS2000}%
  \BibitemOpen
  \bibfield  {author} {\bibinfo {author} {\bibfnamefont {M.}~\bibnamefont
  {{Vojta}}}, \bibinfo {author} {\bibfnamefont {C.}~\bibnamefont
  {{Buragohain}}}, \ and\ \bibinfo {author} {\bibfnamefont {S.}~\bibnamefont
  {{Sachdev}}},\ }\bibfield  {title} {\enquote {\bibinfo {title} {{Quantum
  impurity dynamics in two-dimensional antiferromagnets and
  superconductors}},}\ }\href {\doibase 10.1103/PhysRevB.61.15152} {\bibfield
  {journal} {\bibinfo  {journal} {Phys. Rev. B}\ }\textbf {\bibinfo {volume}
  {61}},\ \bibinfo {pages} {15152} (\bibinfo {year} {2000})},\ \Eprint
  {http://arxiv.org/abs/cond-mat/9912020} {arXiv:cond-mat/9912020
  [cond-mat.str-el]} \BibitemShut {NoStop}%
\bibitem [{\citenamefont {{Sachdev}}(2001)}]{SS2001}%
  \BibitemOpen
  \bibfield  {author} {\bibinfo {author} {\bibfnamefont {S.}~\bibnamefont
  {{Sachdev}}},\ }\bibfield  {title} {\enquote {\bibinfo {title} {{Static hole
  in a critical antiferromagnet: field-theoretic renormalization group}},}\
  }\href {\doibase 10.1016/S0921-4534(01)00198-8} {\bibfield  {journal}
  {\bibinfo  {journal} {Physica C Superconductivity}\ }\textbf {\bibinfo
  {volume} {357}},\ \bibinfo {pages} {78} (\bibinfo {year} {2001})},\ \Eprint
  {http://arxiv.org/abs/cond-mat/0011233} {arXiv:cond-mat/0011233
  [cond-mat.str-el]} \BibitemShut {NoStop}%
\bibitem [{\citenamefont {{Vojta}}\ and\ \citenamefont
  {{Fritz}}(2004)}]{VojtaFritz04}%
  \BibitemOpen
  \bibfield  {author} {\bibinfo {author} {\bibfnamefont {M.}~\bibnamefont
  {{Vojta}}}\ and\ \bibinfo {author} {\bibfnamefont {L.}~\bibnamefont
  {{Fritz}}},\ }\bibfield  {title} {\enquote {\bibinfo {title} {{Upper critical
  dimension in a quantum impurity model: Critical theory of the asymmetric
  pseudogap Kondo problem}},}\ }\href {\doibase 10.1103/PhysRevB.70.094502}
  {\bibfield  {journal} {\bibinfo  {journal} {Phys. Rev. B}\ }\textbf {\bibinfo
  {volume} {70}},\ \bibinfo {eid} {094502} (\bibinfo {year} {2004})},\ \Eprint
  {http://arxiv.org/abs/cond-mat/0309262} {arXiv:cond-mat/0309262
  [cond-mat.str-el]} \BibitemShut {NoStop}%
\bibitem [{\citenamefont {{Fritz}}\ and\ \citenamefont
  {{Vojta}}(2004)}]{FritzVojta04}%
  \BibitemOpen
  \bibfield  {author} {\bibinfo {author} {\bibfnamefont {L.}~\bibnamefont
  {{Fritz}}}\ and\ \bibinfo {author} {\bibfnamefont {M.}~\bibnamefont
  {{Vojta}}},\ }\bibfield  {title} {\enquote {\bibinfo {title} {{Phase
  transitions in the pseudogap Anderson and Kondo models:{\quad} Critical
  dimensions, renormalization group, and local-moment criticality}},}\ }\href
  {\doibase 10.1103/PhysRevB.70.214427} {\bibfield  {journal} {\bibinfo
  {journal} {Phys. Rev. B}\ }\textbf {\bibinfo {volume} {70}},\ \bibinfo {eid}
  {214427} (\bibinfo {year} {2004})},\ \Eprint
  {http://arxiv.org/abs/cond-mat/0408543} {arXiv:cond-mat/0408543
  [cond-mat.str-el]} \BibitemShut {NoStop}%
\bibitem [{\citenamefont {{Fritz}}(2006)}]{FritzThesis}%
  \BibitemOpen
  \bibfield  {author} {\bibinfo {author} {\bibfnamefont {L.}~\bibnamefont
  {{Fritz}}},\ }\bibfield  {title} {\enquote {\bibinfo {title} {{Quantum Phase
  Transitions in Models of Magnetic Impurities}},}\ }\href@noop {} {\bibfield
  {journal} {\bibinfo  {journal} {Ph. D. Dissertation, Universit\"at
  Karlsruhe}\ } (\bibinfo {year} {2006})}\BibitemShut {NoStop}%
\bibitem [{\citenamefont {Si}\ and\ \citenamefont {Kotliar}(1993)}]{Si1993a}%
  \BibitemOpen
  \bibfield  {author} {\bibinfo {author} {\bibfnamefont {Q.}~\bibnamefont
  {Si}}\ and\ \bibinfo {author} {\bibfnamefont {G.}~\bibnamefont {Kotliar}},\
  }\bibfield  {title} {\enquote {\bibinfo {title} {{Fermi-liquid and
  non-Fermi-liquid phases of an extended Hubbard model in infinite
  dimensions}},}\ }\href {\doibase 10.1103/PhysRevLett.70.3143} {\bibfield
  {journal} {\bibinfo  {journal} {Phys. Rev. Lett.}\ }\textbf {\bibinfo
  {volume} {70}},\ \bibinfo {pages} {3143} (\bibinfo {year}
  {1993})}\BibitemShut {NoStop}%
\bibitem [{\citenamefont {{Si}}\ and\ \citenamefont
  {{Kotliar}}(1993)}]{Si1993b}%
  \BibitemOpen
  \bibfield  {author} {\bibinfo {author} {\bibfnamefont {Q.}~\bibnamefont
  {{Si}}}\ and\ \bibinfo {author} {\bibfnamefont {G.}~\bibnamefont
  {{Kotliar}}},\ }\bibfield  {title} {\enquote {\bibinfo {title} {{Metallic
  non-Fermi-liquid phases of an extended Hubbard model in infinite
  dimensions}},}\ }\href {\doibase 10.1103/PhysRevB.48.13881} {\bibfield
  {journal} {\bibinfo  {journal} {Phys. Rev. B}\ }\textbf {\bibinfo {volume}
  {48}},\ \bibinfo {pages} {13881} (\bibinfo {year} {1993})},\ \Eprint
  {http://arxiv.org/abs/cond-mat/9307060} {arXiv:cond-mat/9307060 [cond-mat]}
  \BibitemShut {NoStop}%
\bibitem [{\citenamefont {{Sachdev}}\ \emph {et~al.}(1994)\citenamefont
  {{Sachdev}}, \citenamefont {{Senthil}},\ and\ \citenamefont
  {{Shankar}}}]{SSS94}%
  \BibitemOpen
  \bibfield  {author} {\bibinfo {author} {\bibfnamefont {S.}~\bibnamefont
  {{Sachdev}}}, \bibinfo {author} {\bibfnamefont {T.}~\bibnamefont
  {{Senthil}}}, \ and\ \bibinfo {author} {\bibfnamefont {R.}~\bibnamefont
  {{Shankar}}},\ }\bibfield  {title} {\enquote {\bibinfo {title}
  {{Finite-temperature properties of quantum antiferromagnets in a uniform
  magnetic field in one and two dimensions}},}\ }\href {\doibase
  10.1103/PhysRevB.50.258} {\bibfield  {journal} {\bibinfo  {journal} {Phys.
  Rev. B}\ }\textbf {\bibinfo {volume} {50}},\ \bibinfo {pages} {258} (\bibinfo
  {year} {1994})},\ \Eprint {http://arxiv.org/abs/cond-mat/9401040}
  {arXiv:cond-mat/9401040 [cond-mat]} \BibitemShut {NoStop}%
\bibitem [{\citenamefont {{Parcollet}}\ \emph {et~al.}(1998)\citenamefont
  {{Parcollet}}, \citenamefont {{Georges}}, \citenamefont {{Kotliar}},\ and\
  \citenamefont {{Sengupta}}}]{PGKS97}%
  \BibitemOpen
  \bibfield  {author} {\bibinfo {author} {\bibfnamefont {O.}~\bibnamefont
  {{Parcollet}}}, \bibinfo {author} {\bibfnamefont {A.}~\bibnamefont
  {{Georges}}}, \bibinfo {author} {\bibfnamefont {G.}~\bibnamefont
  {{Kotliar}}}, \ and\ \bibinfo {author} {\bibfnamefont {A.}~\bibnamefont
  {{Sengupta}}},\ }\bibfield  {title} {\enquote {\bibinfo {title}
  {{Overscreened multichannel SU(N) Kondo model: Large-N solution and conformal
  field theory}},}\ }\href {\doibase 10.1103/PhysRevB.58.3794} {\bibfield
  {journal} {\bibinfo  {journal} {Phys. Rev. B}\ }\textbf {\bibinfo {volume}
  {58}},\ \bibinfo {pages} {3794} (\bibinfo {year} {1998})},\ \Eprint
  {http://arxiv.org/abs/cond-mat/9711192} {arXiv:cond-mat/9711192
  [cond-mat.str-el]} \BibitemShut {NoStop}%
\bibitem [{\citenamefont {{Parcollet}}\ and\ \citenamefont
  {{Georges}}(1999)}]{PG98}%
  \BibitemOpen
  \bibfield  {author} {\bibinfo {author} {\bibfnamefont {O.}~\bibnamefont
  {{Parcollet}}}\ and\ \bibinfo {author} {\bibfnamefont {A.}~\bibnamefont
  {{Georges}}},\ }\bibfield  {title} {\enquote {\bibinfo {title}
  {{Non-Fermi-liquid regime of a doped Mott insulator}},}\ }\href {\doibase
  10.1103/PhysRevB.59.5341} {\bibfield  {journal} {\bibinfo  {journal} {Phys.
  Rev. B}\ }\textbf {\bibinfo {volume} {59}},\ \bibinfo {pages} {5341}
  (\bibinfo {year} {1999})},\ \Eprint {http://arxiv.org/abs/cond-mat/9806119}
  {cond-mat/9806119} \BibitemShut {NoStop}%
\end{thebibliography}%

\end{document}